\documentclass[aps,prd,eqsecnum,preprint,
amssymb,amsmath]{revtex4}

\usepackage{latexsym}
\usepackage{epsfig}
\usepackage{graphicx,subfigure}
\usepackage{ulem}
\usepackage{color}
\begin{document}
 
\thispagestyle{empty}
 
\title{Primordial Black Holes as 
Generators of Cosmic Structures} 

\author{Bernard Carr}\email[]{B.J.Carr@qmul.ac.uk}
\affiliation{ 
School of Physics and Astronomy, Queen Mary University of London, Mile End Road, London E1 4NS, UK.\\
%\affiliation{
Research Center for the Early Universe, University of Tokyo, Tokyo 113-0033, Japan.}
\author{Joseph Silk}\email[]{joseph.silk@physics.ox.ac.uk}
\affiliation{
Department of Physics and Astronomy, Johns Hopkins University, Baltimore, MD 21218, USA.\\
Institut d’Astrophysique de Paris, UMR 7095 CNRS, Sorbonne Universit{\'e},
98 bis Boulevard Arago, 75014 Paris, France.
}
\date{\today}
\begin{abstract}                
Primordial black holes (PBHs) could provide the dark matter in various mass windows 
below $10^2 M_{\odot}$ and those of $30 M_{\odot}$ might explain the LIGO events. 
PBHs much larger than this might 
have important consequences even if they
provide only a small fraction of the dark matter. In particular, they could generate cosmological structure either individually through the `seed' effect 
%of a single PBH  
or collectively through the `Poisson' effect,
% of many of them, 
thereby alleviating some problems associated with the standard CDM scenario.
 If the PBHs all have a similar mass and 
make a small contribution to the dark matter, then the seed effect dominates on small scales, in which case 
PBHs could generate the supermassive black holes in galactic nuclei 
or even galaxies themselves. 
If they have a similar mass and provide the dark matter, the Poisson effect dominates on all scales and 
the first bound clouds 
 would form 
earlier than in the usual scenario, with interesting observational consequences. If the PBHs have an extended mass spectrum, which is more likely, they could fulfill all three roles -- providing  the dark matter, binding  the first bound clouds and generating galaxies. In this case, 
the galactic mass function naturally 
 has the observed form, with the galaxy mass being simply related to the black hole mass.
 The stochastic gravitational wave background from the PBHs in this scenario would extend continuously from the LIGO frequency to  the LISA frequency, offering a potential goal for  future surveys.\\
%{\bf Keywords:} dark matter, early Universe, galaxies: formation, stars: Population III, quasars; supermassive black holes, gravitational waves 
 %quasars: general, 
  \end{abstract}
%\pacs{04.70.Bw, 97.60.Lf, 95.35.+d}

\maketitle

%\tableofcontents

\section{Introduction}

The standard Cold Dark Matter (CDM) scenario is characterised by two assumptions: the dark matter comprises some form of weakly interacting massive particle (WIMP); and cosmic structures -- from the first bound clouds through galaxies to clusters of galaxies -- form from initial inhomogeneities through a process of hierarchical build-up. However, both these assumption may be questioned. After many decades of searching, there is still no evidence for WIMPs, either from accelerator experiments or from dark matter searches (Di Valentino et al. 2014) and simulations of structure formation in the CDM scenario reveal
several well-known problems on the scale of galaxies, 
including  
 missing satellites, cores versus cusps, too big-to-fail, frequency of ultra-diffuse galaxies and the baryon fraction (Silk 2017a).
Another problem is that some observational anomalies may require the existence of non-linear structures early in the history of the Universe (Dolgov 2016). In particular, it is now known that most galactic nuclei  contain supermassive black holes (SMBHs), with mass extending from around $10^{5}\rm M_\odot$ to $10^{10}\rm M_\odot$ and already in place at high redshift
(Kormendy \& Richstone 1995). These SMBHs are usually assumed to form as a result of dynamical processes after galaxy formation but it may be hard to explain how they could have formed so early in the standard picture, especially in dwarf galaxies (Silk 2017b).

In this paper we point out that 
many of these problems may
be solved by invoking a population of primordial black holes (PBHs) which formed in the early Universe (Carr \& Hawking 1974). This view has also  been advocated by Garcia-Bellido (2017), Clesse \&  Garcia-Bellido (2015) and Garcia-Bellido \& Clesse (2017). For example, there are general arguments that PBHs rather than WIMPs may provide the dark matter.
This is because the density of such black holes is not constrained by the limits on the baryonic density implied by big bang nucleosynthesis (BBNS), so they would be natural CDM candidates. 
Furthermore, this has the advantage that -- unlike the situation for WIMPs or other particle 
candidates -- there is no need to invoke new physics (Frampton 2016). 
The PBH dark matter proposal has been emphasized from the earliest days of PBH research (Chapline 1975; Carr 1975)  but it has become particularly popular recently (Carr, Kuhnel \& Sandstad 2016; Chapline \& Frampton 2016) - especially since the discovery of black hole coalesences by  LIGO (Bird et al. 2016; Clesse \&  Garcia-Bellido 2017b),   although this may only require a small fraction of the dark matter to be in PBHs (Sasaki et al. 2016). 
However, there are only a few permissible mass windows in which PBHs could contribute significantly to the dark matter (Carr et al. 2010, 2017a). The most interesting for present considerations  is the intermediate mass range ($10-100 M_{\odot}$)
but there are also windows in the lunar-mass ($10^{20}-10^{24}$g) and the asteroid-mass  ($10^{16}-10^{17}$g) ranges. 

Most relevant to the considerations of this paper, 
 there are various ways in which sufficiently massive PBHs could affect the development of large-scales structure and thus help resolve the problems of the CDM scenario.
For example, sufficiently large PBHs might grow enough through accretion  to seed the SMBHs which reside in AGN (Bean \& Magueijo 2002; Clesse \& Garcia-Bellido 2015; Habouzit, Volonteri \& Dubrois 2017). Or if the SMBHs are themselves primordial, they might play a role in generating galaxies, either through the Poisson fluctuations in their number density (Meszaros 1975) or on account of their gravitational Coulomb effect (Hoyle \& Narlikar 1966). In the latter case, they would need to have an initial mass of at least $10^6 \rm M_{\odot}$ but their contribution to the dark matter density need only be $10^{-3}$.
Somewhat smaller PBHs could
allow the first baryonic clouds to bind earlier than usual, with important implications for observations in the dark ages, such as the generation of an infrared background (Kashlinsky 2016). 
This could also modify baryonic feedback in dwarf galaxies (Silk 2017c) and have other knock-on effects for the development of cosmic structure. 

All these features apply
within the standard models of particle physics and 
cosmology, so this proposal  should be regarded as complementing the CDM scenario rather than rivalling it. One just needs to invoke extra non-Gaussian power on scales well below those observable in  the CMB or galaxy surveys. 
Indeed, this illustrates an important principle: one expects the first bound objects to be much smaller than galaxies in most cosmological scenarios and - as discussed by Carr and Rees (1984) even before the advent of the CDM scenario  - many astrophysical processes associated with these objects
 could generate larger scale density fluctuations.  
Thus structure on the scale of galaxies and clusters need not derive entirely from primordial fluctuations. 

But is the existence of such huge PBHs plausible? 
A PBH forming  at a time $t$ after the big bang would have a mass of order the particle horizon size $\sim 10^5(t/s)\rm M_{\odot}$, so this depends on how late they  can form.  It is sometimes argued that this
should be before weak freeze-out at $1$~s, corresponding to a maximum mass of  $10^5\rm M_{\odot}$.
This is because PBH production usually requires large inhomogeneities, which might be expected to disturb the usual BBNS scenario.
However, this argument is not clear-cut  because the fraction of the universe in PBHs at a time $t$ after the Big Bang  is only $\sim 10^{-6} \Omega_{\mathrm{PBH}} (t/s)^{1/2}$, where $\Omega_{\mathrm{PBH}}$ is the current PBH density in units of the critical density (Carr 1975), so this would be at most $10^{-6}$ at weak freeze-out. 
Therefore it is not clear that this disturbs BBNS, although 
it does require fine-tuning of the collapse fraction.

Even if the formation of such large PBHs is not precluded, could they be {\it expected} to form? As reviewed by Carr (2006), PBHs may be generated by three mechanisms: through some form of cosmological  phase transition, through a temporary softening of the equation of state or through the collapse of large inhomogeneities.  The first two mechanisms are unlikely to be relevant after $1$~s but the third one could be. For example, hybrid inflation could produce a spike in the power-spectrum of density fluctuations at a mass-scale which is essentially arbitrary (Garcia-Bellido, Linde \& Wands 1996). Indeed, many people have argued for a spike or non-Gaussianity in the intermediate mass range ($10-10^3 M_{\odot}$) in order to explain the dark matter 
with PBHs (Frampton et al. 2010; Byrnes, Copeland \& Green 2010; Motohashi \& Hu 2017; Garcia-Bellido \& Ruiz Morales 2017, 2018). 

The
proposals that the dark matter comprises PBHs and that supermassive PBHs provide seeds for galaxies are essentially independent, since the mass scale  ($m$) and PBH dark matter fraction 
($f$) are very different. One requires $f \sim 1$ and $m < 10^2 \rm M_{\odot}$ for the dark matter but  $f \sim 10^{-3}$ and $m >10^9 \rm M_{\odot}$ for galactic seeds.
Clearly each scenario is of interest in its own right. However, it is important to note that generic initial conditions of PBH formation suggest that their masses
should extend over a wide range (Carr et al. 2016),
so it is possible that they could serve {\it both} functions.  
This means that one could have a significant density of PBHs well above the mass of those which provide the dark matter and possibly as large as  $\sim 10^6\rm M_\odot$. This would also have important implications for the existence of a stochastic gravitational wave background (Clesse \& Garcia-Bellido 2015). 

The plan of this paper is as follows: Sec.~II summarizes constraints on the fraction of the dark matter 
in large PBHs.  Sec.~III reviews previous work on the expected PBH mass function for various scenarios. Sec.~IV discusses the generation of fluctuations by the seed and Poission effect for both a monochromatic and extended PBH mass function, identifying the dominant effect in various  astronomical contexts. Sec.~V derives constraints on the PBH dark matter fraction in order to avoid 
cosmic structures forming too early. Sec.~VI considers whether the SMBHs in galactic nuclei could be primordial and thus seed galaxies, pointing out that sufficiently large PBHs would swallow their host galaxy entirely.  It also considers the effects of PBHs on the formation of the first bounds clouds.
Sec.~VII discusses the possible
gravitational wave background generated by PBHs with an extended mass function. Sec.~VIII draws some general conclusions. 

\section{Constraints on massive PBHs}

\begin{figure}
 \begin{center}
\psfig{file=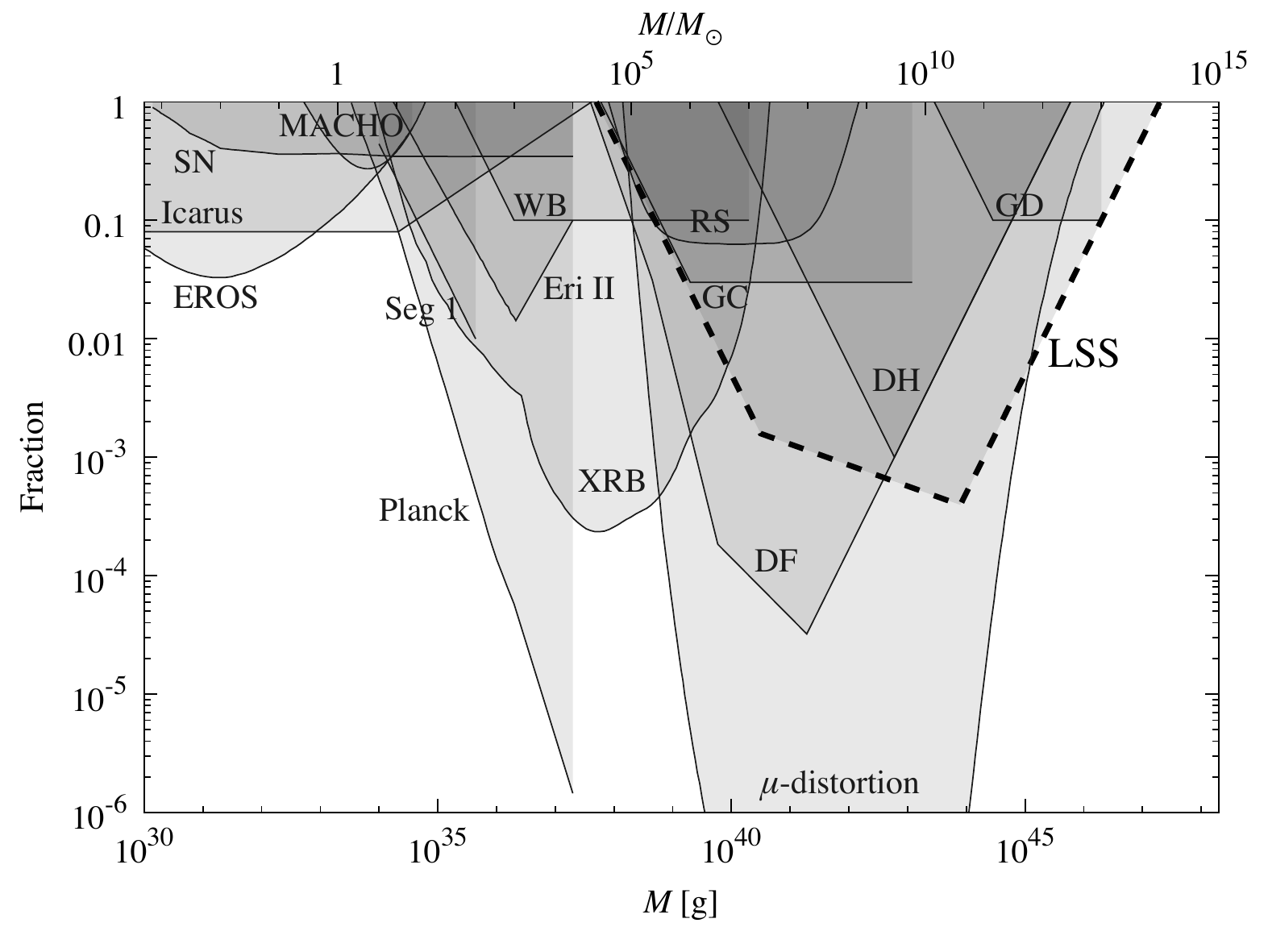,width=4.5in}
\caption {
Constraints on the fraction of the dark matter in large PBHs from a variety of lensing, dynamical and accretion effects, taken from Carr et al. (2018): microlensing of stars in the LMC (MACHO/EROS) and in the giant arcs of cluster lenses (Icarus); microlensing of supernovae (SN);
millilensing of radio sources (RS); disruption of wide binaries (WB), globular clusters (GC) and star clusters in Eridanus (Eri II); disk-heating (DH) and disruption of dwarf galaxy Segue 1 (Seg I); dynamical friction drag of halo objects (DF) and galaxy disruption  in clusters (GD); accretion constraints from  CMB anisotropy (Planck) and X-ray binaries (XRB).
Also shown is  the range of masses excluded by the $\mu$-distortion constraint if PBHs form from Gaussian primordial fluctuations. The large-scale structure limit (derived in Sec.~V) is shown as a broken bold line. 
} 
 \label{limit}
 \end{center}
 \end{figure}

We now briefly review various constraints which can be imposed on the density of PBHs large enough to affect the development of cosmic structures. These constraints have been discussed in many recent works but an up-to-date summary for a monochromatic PBH mass function is shown in Fig.~\ref{limit}. This is 
part of a figure from Carr et al. (2018), which provides a comprehensive  review  of the constraints over a much wider mass range.  
Clearly there is considerable overlap with these constraints but they all come with various caveats, so it is useful to have several in each mass range. While PBHs with a single mass are excluded from providing all the dark matter ($f \sim 1$) over the entire mass range above about $1M_{\odot}$, the main message of Fig.~\ref{limit} is that  this may not exclude them having the small density ($ f \sim 10^{-3}$) required for cosmic structure effects. In particular, the fraction of the dark matter density  in SMBHs in galactic nuclei is $\Omega_{smbh}/ \Omega_{dm} \sim   10^{-4}$.
Perhaps the most serious constraint  comes from the $\mu$-distortion  in the CMB, expected if PBHs are generated by primordial inhomogeneities on scales which are later dissipated.  This may exclude PBHs in the mass range $10^5 - 10^{12}M_{\odot}$ entirely, so we give this special consideration.

Lensing effects constrain $f(m)$ over a wide range of masses. The microlensing of stars in the LMC by objects in our own halo was studied by  
the MACHO and EROS experiments (Alcock et al. 2001, Hamadache et al. 2006) and provides constraints up to around $10 M_{\odot}$, although these constraints may be weakened  for certain halo models (Calcino, Garcia-Bellido \& Davis 2018). 
Studies of quasar microlensing by Mediavilla et al. (2009) suggest a limit $f(m) < 0.05$ for $0.1\,M_{\odot}< m < 10\,M_{\odot}$ (i.e. the same mass rnage)  but the microlensing of supernovae (Zumalacarregui \& Seljak 2017) or stars in the giant arcs gererated by cluster lensing (Oguri et al. 2018)  extends the constraints to masses of around $10^4 M_{\odot}$. 
At still larger mass scales, searches for millilensing of compact radio sources by Wilkinson et al. (2001) give a limit
\begin{equation}
	f( m )
		<
				\begin{cases}
						( m / 2 \times 10^{4}\, M_{\odot} )^{-2}
					& ( m < 10^{5}\,M_{\odot} )\,  \\
						0.06
					& ( 10^{5}\,M_{\odot}< m < 10^{8}\,M_{\odot} )\,  \\
						(m / 4 \times 10^{8}\,M_{\odot} )^{2}
					& ( m > 10^{8}\,M_{\odot} )\, .
				\end{cases}
\end{equation}
We do not discuss the various claims of positive detections of dark matter by lensing.

Numerous dynamical limits have been discussed  by Carr \& Sakellariadou (1999).
Many of them involve the destruction of various astronomical objects by the passage of nearby PBHs. If the PBHs have density $\rho$ and velocity dispersion $V$, while the objects have mass $M_c$, radius $R_c$, velocity disperson $V_c$ and survival time $t_L$, then the constraint has the form:
\begin{equation}
	f( m) <
				\begin{cases}
						M_c V/(Gm \rho  t_L R_c) 
					& ( m < M_c(V/V_c) )\,  \\
						M_c/(\rho V_c t_L R_C^2)
					& ( M_c(V/V_c) < m < M_c(V/V_c)^3) \  \\
						m V_c^2/( \rho R_c^2 V^3 t_L) \exp [(m/M_c)(V_c/V)^3]
					& ( m > M_c(V/V_c)^3) \, .
				\end{cases}
\end{equation}
The three limits correspond to disruption by multiple encounters, one-off 
encounters and non-impulsive encounters, respectively. The fraction is thus constrained over the mass range
\begin{equation}
 \frac{M_c V}{G \rho_{\rm dm} t_L R_c} < m < 
 M_c \left( \frac{V}{V_c}\right)^3 \, ,
\end{equation}
the limits corresponding to the values of $m$  for which one could have $f \sim 1$. 
Various numerical factors 
in the analysis of Carr \& Sakellariadou are omitted in this  discussion. 

This argument can be applied at low $m$ to wide binaries in the Galactic disk, 
these being especially vulnerable to disruption from PBH encounters.
By comparing the result of simulations with observations, Yoo, Chaname \& Gould (2004)  ruled out objects with $ m > 43\,M_\odot $ from providing most of the halo mass.
Later Quinn et al. (2009) argued
that one of the widest-separation binaries was spurious, leading to the weaker constraint  $ m > 500\,M_\odot $.
However, the most recent analysis comes from Monroy-Rodriguez \& Allen (2014) and may reduce the limiting mass to around $20 M_{\odot}$. As a compromise, we take the limit to be  $100 M_{\odot}$ and the constraint  then becomes
\begin{equation}
f(m)
<
\begin{cases}
(m/10^2\,M_\odot)^{-1}
& (10^2\,M_\odot < m < 10^3\,M_\odot )\,  \\
0.1
& (10^3 M_\odot < m <10^7 M_\odot)\, . 
\end{cases}
\label{wb}
\end{equation}
Using a similar argument, 
the survival of globular clusters against tidal disruption by halo PBHs gives a limit (Moore 1993; Carr \& Sakellariadou 1999)
\begin{equation}
	f( m )
		<
				\begin{cases}
						( m / 3 \times 10^{4}\,M_{\odot} )^{-1}
					& ( 3 \times 10^{4}\,M_{\odot} < m < 10^{6}\,M_{\odot} )\,  \\
						0.03
					& ( 10^{6}\,M_{\odot} < m < 10^{9}\,M_{\odot} )\, , 
				\end{cases}
\end{equation}
although this depends on the mass and the radius of the typical cluster. 
Related but somewhat  stronger  constraints are associated with the survival of Segue I (Koushiappas \& Loeb 2017)  and a star cluster in the dwarf galaxy Eridanus II (Brandt 2016) but
we note that the presence of a primordial IMBH in an ultra-faint dwarf would enhance its survivability. On still larger scales the survival of galaxies in clusters against disruption by giant cluster PBHs gives a limit (Carr \& Sakellariadou 1999)
\begin{equation}
	f( m )
		<
				\begin{cases}
						( m / 10^{10}\,M_{\odot} )^{-1}
					& (10^{10}\,M_{\odot} < m < 10^{11}\,M_{\odot} )\,  \\
						0.1
					& ( 10^{11}\,M_{\odot} < m < 10^{13}\,M_{\odot} )\, . 
				\end{cases}
\end{equation}

A somewhat different dynamical effect is the heating of stars in the Galactic disc by halo objects, this giving a limit 
\begin{equation}
	f( m )
		<
				\begin{cases} 
						( m / 3 \times 10^{6}\,M_{\odot} )^{-1}
					& (3 \times 10^{6}\,M_{\odot} < m < 3 \times 10^{9}\,M_{\odot} ) \,  \\
						( m / M_{\mathrm{halo}} )
					& (  3 \times 10^{9}\,M_{\odot} < m < M_{\mathrm{halo}} \sim 10^{12}\,M_{\odot}) \, .
				\end{cases}
				\label{disc}
\end{equation}
The lower expression corresponds to having at least one PBH within the Galactic halo.
The upper limit of $ 3 \times 10^{6}\,M_{\odot}$ agrees with the more precise calculations by Lacey \& Ostriker (1985), although they argued that black holes with $ 2 \times 10^{6}\,M_{\odot}$ could {\it explain} some features of disc heating. Constraint \eqref{disc} bottoms out
with a value $f \sim 10^{-3}$. On still larger scales, dynamical friction will drag halo black holes into the Galactic nucleus;  these holes could then merge to form a single SMBH whose mass must not exceed the observed SMBH mass of $4 \times 10^6 M_{\odot}$. We include this limit in  Fig.~\ref{limit} but do not give an explicit expression for it  since it is complicated and 
model-dependent.
Also it can be circumvented if black holes can be ejected from the Galactic nucleus by 3-body effects. 
\if
The limit applies providing there is at least one PBH within the relevant scale (i.e. the Galactic halo for halo dark matter), which corresponds to the ``incredulity'' limit 
\begin{equation}
	f( m)
		>
			( m / M_{\mathrm{halo}} ),
			\quad
	M_{\mathrm{halo}}
		\approx
			3 \times 10^{12}\,M_{\odot}
			\, .
			\label{incredulity}
\end{equation}
\fi

Another important constraint comes from accretion effects.
PBHs cannot accrete appreciably in the radiation-dominated era {(Carr \& Hawking 1974; Novikov et al. 1979) but 
they might still do so in the matter-dominated period after decoupling and a Bondi-type analysis should then apply. The associated accretion and emission of radiation could have an important effect on the thermal history of the Universe, as first analysed by Carr (1981). This possibility was investigated in more detail by Ricotti, Ostriker \& Mack (2008), who studied the effects of such accreting PBHs on the ionisation and temperature evolution of the Universe. The emitted X-rays would produce anisotropies and spectral distortions in the cosmic microwave background (CMB). Using WMAP data to constrain the first, they obtained the constraint:
\begin{equation}
	f( m )
		<
				\begin{cases}
						( m / 30\,M_{\odot} )^{-2}
					& ( 30\,M_{\odot}< m < 10^{4}\,M_{\odot} )\, ,\\
						10^{-5}
					& ( 10^{4}\,M_{\odot} < m < 10^{11}\,M_{\odot} )\, .\\
				\end{cases}
\label{wmap}
\end{equation}
The limit flattens off above $10^4M_{\odot}$ because the black hole acretion rate then exceeds the Eddington limit. 
The spectral distortion limit implied by FIRAS data has a similar form but extends down to a lower mass and
bottoms out a larger value of $f$. 

The Ricotti et al. limit is not  shown in Fig.~\ref{limit}  because it contains an error.  However,  
recently the accretion constraints have been reconsidered by several  groups, who 
find weaker limits. Ali-Haimoud \& Kamionkowski (2016) calculate the accretion on the assumption that it is suppressed by Compton drag and Compton cooling from CMB photons, allowing for the PBH velocity relative to the background gas. They find the spectral distortions are too small to be detected, 
while the anisotropy constraints 
only exclude $f=1$ above $10^2 M_{\odot}$.
Horowitz (2016) and Chen, Huang \& Wang (2016) perform a similar analysis and obtain stronger upper limits of $30 M_{\odot}$ and $0.2 M_{\odot}$, respectively.
However, neither of these analyses includes 
the  flattening of the limit on $f(m)$ above some mass due to the accretion rate exceeding the Eddington limit. 
The CMB anisotropy constraints from the Planck satellite are even stronger if the PBHs form accretion discs (Poulin  et al. 2017). 
This is the only accretion limit shown in Fig.~\ref{limit} but we stress that all of them are very dependent on astrophysical assumptions and therefore not so secure. We also show the limit associated with X-ray source counts (Inoue \& Kusenko2017), this extending from a few to $10^7 M_{\odot}$. 
 
Another important limit 
comes from the dissipation of density fluctuations between  $10^6$ and $10^9$s (or $5\times 10^4<z<2\times 10^6$) by Silk damping. This results in a $\mu$-distortion in the CMB spectrum (Chluba, Erickcek \& Ben-Dayan 2012), leading to an upper limit $\delta (m) < \sqrt{\mu} \sim 10^{-2}$ over the mass range $10^3 < m/M_{\odot} < 10^{12}$. This limit was first given in Carr \& Lidsey (1993), based on a result in Barrow \& Coles (1991), but the limit on $\mu$  is now  stronger. 
When PBHs form abundantly, the dispersion of primordial fluctuations is also expected to be large, so Silk damping would produce unacceptably large $\mu$ distortions.
However, this is  a limit on  the density fluctuations from which the PBHs derive and it can be translated into an upper limit on the PBH abundance only if one assumes a model for their formation. 

If the fluctuations are Gaussian and the PBHs form on the  high-$\sigma$ tail, as in the simplest scenario (Carr 1975), one finds a constraint on $f(m)$ in the range $10^3 < m/M_{\odot} < 10^{12}$
(Kohri et al. 2014). 
However, the assumption that the PBHs form on the high-$\sigma$ tail of Gaussian density fluctuations 
may be incorrect. 
For example, this does not apply in the model of Garcia-Bellido, Peloso \& Unal (2017) or in the ``patch'' model of Nakama, Suyama \& Yokoyama (2016), in which the relationship between the background inhomogeneities 
and the overdensity in the tiny fraction of the volume which collapses to PBHs is modified.  
The $\mu$-distortion constraint could thus be much weaker, so one  
needs to consider the dependence of  the $\mu-$distortion limits on the possible non-Gaussianity of primordial fluctuations. 

By using a phenomenological description of non-Gaussianity introduced in Nakama, Suyama \& Yokoyama (2014), Nakama, Carr \& Silk (2018) have recently calculated the constraints on $f(m)$, assuming both the FIRAS limit of $\mu=9 \times 10^{-5}$ (Planck collaboration 2016) and  the projected upper limit of $\mu<3.6\times 10^{-7}$ from PIXIE (Abitbol et al. 2017).
%They use Eq.~(7) of Nakama, Silk \& Kamionkowski  (2017b) to convert PBH mass to wave-number and Eq.~(20)  to convert $\beta$ to $f$.
%=\Omega_{\mathrm{PBH}}/\Omega_{\mathrm{dm}}$.
The limits are shown by the grey band in Fig.~\ref{limit} and essentially rule out PBHs playing an important cosmological role over the entire mass range $10^5 - 10^{12} M_{\odot}$ unless the primordial fluctuations are highly non-Gaussian. 
Otherwise one would need to  invoke smaller PBHs with initial masses of $10^5M_\odot$ which undergo substantial accretion between the $\mu$-disortion era and the time of matter-radiation equality,

\section{The PBH mass function} 

In many scenarios, one would expect PBHs to form with an extended mass function. This is interesting because it would allow them to play a variety of cosmological roles. In this section, we discuss four such scenarios, with particular regard to the question of whether PBHs could  provide both the dark matter and the seeds for cosmic structure. The first assumes that the PBHs form from 
scale-invariant primordial fluctuations or the collapse of cosmic strings, the second that they  form in an early matter-dominated era, the third  that they form from initial inhomogeneities of inflationary origin, and the  
 fourth that they form from critical collapse. In each of these cases, we will give the form of the mass function and  the relative densities of the PBHs which provide the dark matter and the cosmic seeds. If the SMBHs in galactic nuclei are primordial, observations require the ratio of the densities to be of order $10^{-4}$ but the initial ratio may be smaller if the PBHs accrete. 

\subsection{Collapse from scale-invariant fluctuations or cosmic strimgs}

If the PBHs form from scale-invariant fluctuations (i.e. with constant amplitude at the horizon epoch),
their mass spectrum should have the power-law form (Carr 1975)
\begin{equation}
\frac{dn}{dm} \propto m^{-\alpha} \quad \mathrm{with} \quad  \alpha = \frac {2(1+ 2 \gamma)}{1+ \gamma}   \, ,
\label{spectrum}
\end{equation}
where $\gamma$ specifies the equation of state ($p = \gamma \rho c^2$) at PBH formation.  The exponent arises because the background density and PBH density have different redshift dependencies. The mass function is also proportional to the probability $\beta$ that an overdense region of mass $m$ has a size exceeding the Jeans length at maximum expansion, so that it can collapse against the pressure. In this case, $\beta$ should be scale-independent, so if the horizon-scale fluctuations have a Gaussian distribution with dispersion $\delta_H$, one expects (Carr 1975)
\begin{equation}
\beta \approx \rm{erfc} \,  (\delta_c/ \sqrt{2} \delta_H) \, .
\label{beta}
\end{equation}
Here erfc is the complimentary error function and $\delta_c$ is the threshold for PBH formation.  A simple analytic argument suggest $\delta_c \approx \gamma$ but more precise arguments -- both numerical (Musco \& Miller 2013)   and analytical (Harada, Yoo \& Kohri 2013) -- suggest a somewhat larger value.  
At one time it was argued that the primordial fluctuations would be {\it expected} to be scale-invariant (Harrison 1971) but this does not apply in the inflationary scenarios (discussed below). Nevertheless, one would still expect the above equations to apply if the PBHs form from the collapse of cosmic loops because the collapse probability is then scale-invariant. 

One usually assumes $0 < \gamma <1$, corresponding to $2 < \alpha < 3$, in which case most of the density is in the smallest PBHs and
the density of those larger than $m$ is 
\begin{equation}
 \rho(m) = \int^{m_{max}}_{m} m (dn/dm) dm \propto m^{2-\alpha}  \quad (m_{min}<m < m_{max}) \, ,
\end{equation} 
where $m_{max}$ and $m_{min}$ are the upper and lower cut-offs for the mass function. 
If we assume that  the  PBHs contain a fraction $f_{dm~}$ of the dark matter, this implies
that the fraction of the dark matter  in PBHs of mass larger than $m$ is  
\begin{equation}
f(m) \equiv \rho(m)/ \rho_{dm} \approx f_{dm} (m_{dm}/m)^{\alpha -2}   \quad (m_{min} < m < m_{max}) \, ,
\label{dark}
\end{equation}
where $m_{dm} \approx m_{min}$ is the mass-scale which contains most of the dark matter.
[Alternatively, one could define $f(m)$ as the fraction in PBHs in the mass interval ($m,2m$), which is smaller by a factor $1-2^{2 - \alpha}$.] In a radiation-dominated era, which is most likely, $\gamma =1/3$ and the exponent in Eq.~(\ref{dark}) becomes $1/2$. There is then a simple relationship between the density of the primordial SMBHs, 
taken to have a 
mass $m_{smbh}$, and ones which provide the dark matter: 
\begin{equation}
f_{smbh}/f_{dm} \sim (m_{dm}/m_{smbh})^{1/2} \sim 10^{-4} (m_{dm}/10 M_{\odot})^{1/2} (m_{smbh}/10^9 M_{\odot})^{-1/2}  \, .
\end{equation}
If one wants to identify the SMBHs with those in galactic nuclei, this 
ratio must be around $10^{-4}$, which 
requires $m_{smbh} \sim 10^8m_{dm}$. In a more general scenario, in which $\alpha$ is regarded as a free parameter, unrelated to $\gamma$, one requires $m_{smbh} \sim 10^{4/(\alpha -2)} m_{dm}$. 

\subsection{Collapse in a matter-dominated era}

PBHs form more easily if the Universe  becomes pressureless (i.e. matter-dominated) for some period. For example,  this may arise due to some form of phase transition in which the mass is  channeled into non-relativistic particles (Khlopov \& Polnarev 1980;  Polnarev \& Khlopov 1985) or due to slow reheating after inflation (Khlopov, Malomed \& Zeldovich 1985; Carr, Gilbert \& Lidsey 1994; Carr et al. 2018b). Since the value of $\alpha$ in the above analysis is $2$ for $\gamma =0$, one might expect $\rho(m)$ to increases logarithmically with $m$. However, the analysis breaks down in this case because the Jeans length is much smaller than the particle horizon, so pressure does not inhibit collapse at all. Instead, collapse is prevented by  deviations from spherical symmetry and the probabiity of PBH formation can be shown to be (Khlopov \& Polnarev 1980; Polnarev \& Khlopov 1985)
\begin{equation}
\beta(m) = 0.02 \, \delta_H(m)^{5} \, .
 \end{equation}
This is in agreement with the  recent analysis of Harada et al (2016) and  leads to a mass function
\begin{equation}
\frac{dn}{dm} \propto m^{-2} \delta_H(m)^5  \, .
\end{equation}
$\beta(m)$ is small for $\delta_H(m) \ll 1$ but much larger than the exponentially suppressed fraction in the radiation-dominated case. If the matter-dominated phase extends from $t_1$ to $t_2$, PBH formation is enhanced over the mass range 
\begin{equation}
m_{min} \sim M_H(t_1) < m < m_{max} \sim M_H(t_2) \delta_H(m_{max})^{3/2} \, .
\end{equation} 
The lower limit is the horizon mass at the start of matter-dominance and the upper limit is the horizon mass at the epoch when the region which binds at the end of matter-dominance enters the horizon.  This scenario has recently been studied by Carr, Tenkanen \& Vaskonen (2017b).

Since the  primordial fluctuations must be approximately scale-invariant (even in the inflationary scenario), $\beta(m)$ is nearly constant, so  Eq.~(\ref{spectrum}) applies with $\alpha \approx 2$. Thus the mass function is uniquely determined by the values of $t_1$ and $t_2$. Although it could well be extended enough to incorporate  both dark matter and SMBH scales, $f(m)$ should only have a weak dependence on $m$,
so  the ratio $f_{smbh}/f_{dm}$ woud be too large for PBHs to fulfill both roles.
However, we note that in the PBH scenario advocated by Deng \& Vilenkin (2017), one expects a combination of mass functions of the form (\ref{spectrum}), with $\alpha =2$ below some critical mass and  $\alpha =5/2$ above it.

\subsection {Collapse from inflationary fluctuations} 

If the fluctuations generated by inflation have a blue spectrum (i.e. decrease with increasing scale) and  the PBHs form from the high-$\sigma$ tail of the fluctuation distribution, then the exponential factor in Eq.~(\ref{beta}) might  suggest that the PBH mass function should have an exponential upper cut-off at the horizon mass when inflation ends
(Carr et al. 1994). This 
corresponds to the reheat time $t_R$, which
 the CMB quadrupole anisotropy requires to exceed $10^{-35}$s. 
In this case, $f(m)$ should fall off exponentially above the reheat horizon mass, precluding any possibility of PBHs providing both dark matter and SMBHs.
However, a more careful analysis gives a different result. If the fluctuations result from a smooth 
symmetric peak in the inflationary power spectrum (cf. Garcia-Bellido et al. 1996), 
the PBH mass function should have
the lognormal form
\begin{equation}
\frac{dn}{dm} \propto \frac{1}{m^2} \exp \left[ - \frac{(\log m - \log m_c)^2}{2 \sigma^2} \right] \, .
\label{mf}
\end{equation}
%where  $\sigma$ is the dispersion of $\delta$.
This was 
first suggested by Dolgov \& Silk (1993) and later by Clesse \& Garcia-Bellido (2015). It has been demonstrated both numerically (Green 2016) and 
analytically (Kannike et al. 2017) for the case in which the 
slow-roll approximation holds. It is therefore representative of a large class of inflationary scenarios, including 
the axion-curvaton and running-mass infation models considered by Carr et al. (2016).  

Equation~(\ref{mf}) implies that the mass function is symmetric about its peak at $m_c$ and described by two parameters: the mass-scale $m_c$ itself and the width of the distribution $\sigma$.  
The integrated mass function is
\begin{equation}
f(m) = \int_m m \frac{dn}{dm} dm  \approx \rm{erfc} \,  (\ln m/\sigma) \, .
\end{equation}
As in the first scenario, this can explain the dark matter and galactic seeds fairly naturally and it does not require such a broad spread of masses. 
However, not all inflationary scenarios produce the mass function (\ref{mf}). Inomata et al. (2017) propose a scenario which combines a broad mass function at low $m$ (to explain the dark matter) with a sharp one at high $m$ (to explain the LIGO events). On the other hand, one could also envisage a scenario in which the sharp peak is in the SMBH range.

\subsection{Critical collapse}

It is well known that  black hole formation
is associated with critical phenomena (Choptuik 1983)  
and the application of this to PBH formation was first  studied by various authors
(Evans \& Coleman 1994; Koike, Hara \& Adachi 1995; Niemeyer \& Jedamzik 1997). The conclusion was
that the mass function still has an upper cut-off at around the horizon mass 
but there is also a low-mass tail (Yokoyama 1998).
If we assume for simplicity that the density fluctuations have a monochromatic power spectrum
on some mass scale $K$ and identify the amplitude of the density
fluctuation when that scale crosses the horizon, $\delta$, as
the control parameter, then the black hole mass is (Choptuik 1983)
\begin{equation}
m = K\,(\delta - \delta_\mathrm c)^c\, .
\label{eq:2}
\end{equation}
Here $\delta_\mathrm c$ is the critical 
density fluctuation required for PBH formation ($0.4$ in a radiation-dominated era),
the exponent has a universal value  
$c \approx 0.35$ and $K \approx M_H$.
Although the scaling relation \eqref{eq:2} is expected
to be valid only in the immediate neighborhood of
$\delta_\mathrm c$, most black holes should form from fluctuations with this value
because the probability distribution function (PDF)
declines exponentially beyond $\delta=\delta_\mathrm c$ if the fluctuations are blue. 
Hence it is sensible to calculate the expected mass function of
PBHs using Eq.~\eqref{eq:2}.
 This allows us to estimate the mass function
 independently of the specific form of the PDF of primordial
density 
 fluctuations.
A detailed calculation gives the mass function (Yokoyama 1998)
\begin{equation}
\frac{\mathrm dn}{\mathrm dm} \propto
\left(\frac{m}{\gamma\,M_\mathrm f}\right)^{1/c-1}\,
  \exp\left[-(1-c)\,\left(\frac{m}{\gamma\,M_\mathrm f}\right)^{1/c}\right]\, ,
\label{initialMF}
\end{equation}
where
\begin{equation}
 \gamma \equiv \left( \frac{1-c}{s} \right)^c\,, \quad s = \delta_c/\sigma \, , \quad M_f = K \, 
\end{equation}
and $\sigma$ is the dispersion of $\delta$. 
For $c =0.35$, this gives
\begin{equation}
\frac{\mathrm dn}{\mathrm dm} \propto m^{1.85} \exp [-s (m/M_f)^{2.85}] \, .
\end{equation}
The function $f(m)$ would have a similar form but with an exponent of $3.85$ in the first term. In this case, the PBH density is too concentrated around a single mass to produce both the dark matter and the SMBHs in galactic nuclei. 
However, the above analysis depends on the
assumption that the power spectrum of the primordial fluctuations is 
monochromatic. As shown by Kuhnel, Rampf \& Sandstad (2016) 
for a variety of inflationary models, 
when a realistic model of the power
spectrum underlying PBH production is used, the inclusion
 of critical collapse can lead to a significant shift, lowering
and broadening of the PBH mass spectra -- sometimes by
several orders of magnitude. Nevertheless, it still seems unlikely that the PBHs can play more than one role. 

\section{Seed versus Poisson fluctuations}
\label{sec4}

PBHs of mass $m$ provide a source of fluctuations for objects of mass $M$ in two ways: (1) via the seed effect, in which the Coulomb effect of a {\it single} black hole generates an initial density fluctuation $m/M$;
 (2) via the Poisson effect, in which the $\sqrt{N}$ fluctuation in the number of black holes generates an initial density fluctuation $(fm/M)^{1/2}$. Both types of fluctuations then grow through gravitational instability to bind  regions of mass $M$.  
Each of these proposals has a long history, 
although  the early literature tends not to be cited in recent work. The seed mechanism was first proposed by Hoyle and Narlikar (1966) in the context of the Steady State model and later
by Ryan (1972) and Carr \& Rees (1984).
The Poisson mechanism was first suggested by Meszaros (1975), although 
he overestimated the effect 
(Carr 1977), and  it was then explored in many subsequent papers (Freese, Price \& Schramm 1983; Carr \& Silk 1983; Afshordi, McDonald \& Spergel  2003; Chisholm 2006; Kashlinksy 2016). The relationship between the two mechanisms is  subtle, so we will consider both of them in the following discussion and determine the dominant one for each mass scale. 
We first assume that the PBHs have a monochromatic mass function and then consider the effect of an extended mass function, which is more plausible. Note that 
the seed need not be a black hole; a bound cluster of smaller objects (Carr \& Lacey 1987; Metcalf \& Silk 1996)  or Ultra Compact Mini Halos (UCMHs) (Ricotti \& Gould 2009) would serve equally well. Indeed, the density fluctuations required to form UCMHs would be much smaller, so they would generally be more numerous than PBHs (Kohri et al. 2014). 

\subsection{Monochromatic PBH mass function}

If the PBHs have a single mass $m$, 
the initial fluctuation in the matter density on a scale $M$ is
\begin{equation}
\delta_i \approx
\begin{cases}
m/M
& (\mathrm{seed})  \\        
(f m/M)^{1/2}
& (\mathrm{Poisson}) \, ,
\end{cases}
\label{initial}
 \end{equation}
where $f$ is the fraction of the dark matter in the PBHs and $M$ excludes the radiation content. If PBHs provide the dark matter, $f \sim 1$ and the Poisson effect dominates for all $M$ but we also consider scenarios with $f \ll 1$. The Poisson effect then dominates for $M > m/f$ and the seed effect for $M < m/f$. Indeed, the first expression in (\ref{initial}) only applies  for $f < m/M$,
since otherwise a region of mass $M$ would be expected to contain more than one black hole, i.e. 
the mass bound by a single seed can never exceed $m/f$ because of competition from other seeds. 
The dependence of $\delta_i$ on the ratio $M/m$ is indicated in Fig.~\ref{fig2}(a).

It should be stressed that the $\sqrt{N}$ fluctuation does not initially correspond to a fluctuation in the {\it total} density because at formation each PBH is surrounded by a region which is underdense in its radiation density. 
However, because the radiation density falls off faster than the black hole density, a fluctuation in the total density does eventually develop and this has amplitude $\delta_i$ at the horizon epoch. 
(The error in Meszaros's initial analysis was to assume growth of the PBH fluctuation even before then.) Thereafter
one can show (Meszaros 1974) that 
the fluctuation evolves as  
\begin{equation}
\delta = \delta_H \left(1+ \frac{3 \rho_B (t)}{2 \rho_r (t)}\right) \left(1+ \frac{3 \rho_B (t_H)}{2 \rho_r (t_H)}\right)^{-1} \, ,
\label{start}
\end{equation}
where $\rho_B$ and $\rho_r$ are the mean black hole and radiation densities, respectively. Therefore
the $\sqrt{N}$ fluctuation is frozen during the radiation-dominated era but it 
starts growing as $(z+1) ^{-1}$ from  the start of the matter-dominated era ($t > t_{eq}$). [For $f \ll 1$, eqn (\ref{start}) suggests that the PBH fluctuation does not start to grow until $f^{-1} t_{eq}$. However, the associated  fluctuation in the total matter density grows by a factor of $f$ between $t_{eq}$ and $f^{-1} t_{eq}$, so the PBH fluctuation growth effectively starts at $t_{eq}$ even for $f \ll 1$.] 

Since matter-radiation equality  corresponds to a redshift 
$z_{eq} \approx 4000$ and an overdense region binds when $\delta \approx 1$, the mass binding at redshift $z_B$ is 
\begin{equation}
M \approx
\begin{cases}
4000 \, m z_B^{-1}
& (\mathrm{seed})  \\        
10^7 f m z_B^{-2}
& (\mathrm{Poisson}) \, ,
\end{cases}
\label{bind}
 \end{equation}
as illustrated in Fig.~\ref{fig2}(b). Note that one also expects  the peculiar velocity of the PBHs to induce Poisson fluctuations on the scales they can traverse in a cosmological time (Carr \& Rees 1984). In this context, Meszaros considers fluctuations of the form $\Delta N \sim N^{1/3}$, on the assumption that this corresponds to a situation in which the black holes are distributed on a lattice, with their positions being random only on scales smaller than the lattice. However, in this situation it can be shown that the effective fluctuation is really $\Delta N \sim N^{-1/6}$ (Carr 1975) and this is never important. In fact, the above analysis applies even if there are no peculiar velocities.

\begin{figure}
 \begin{center}
\psfig{file=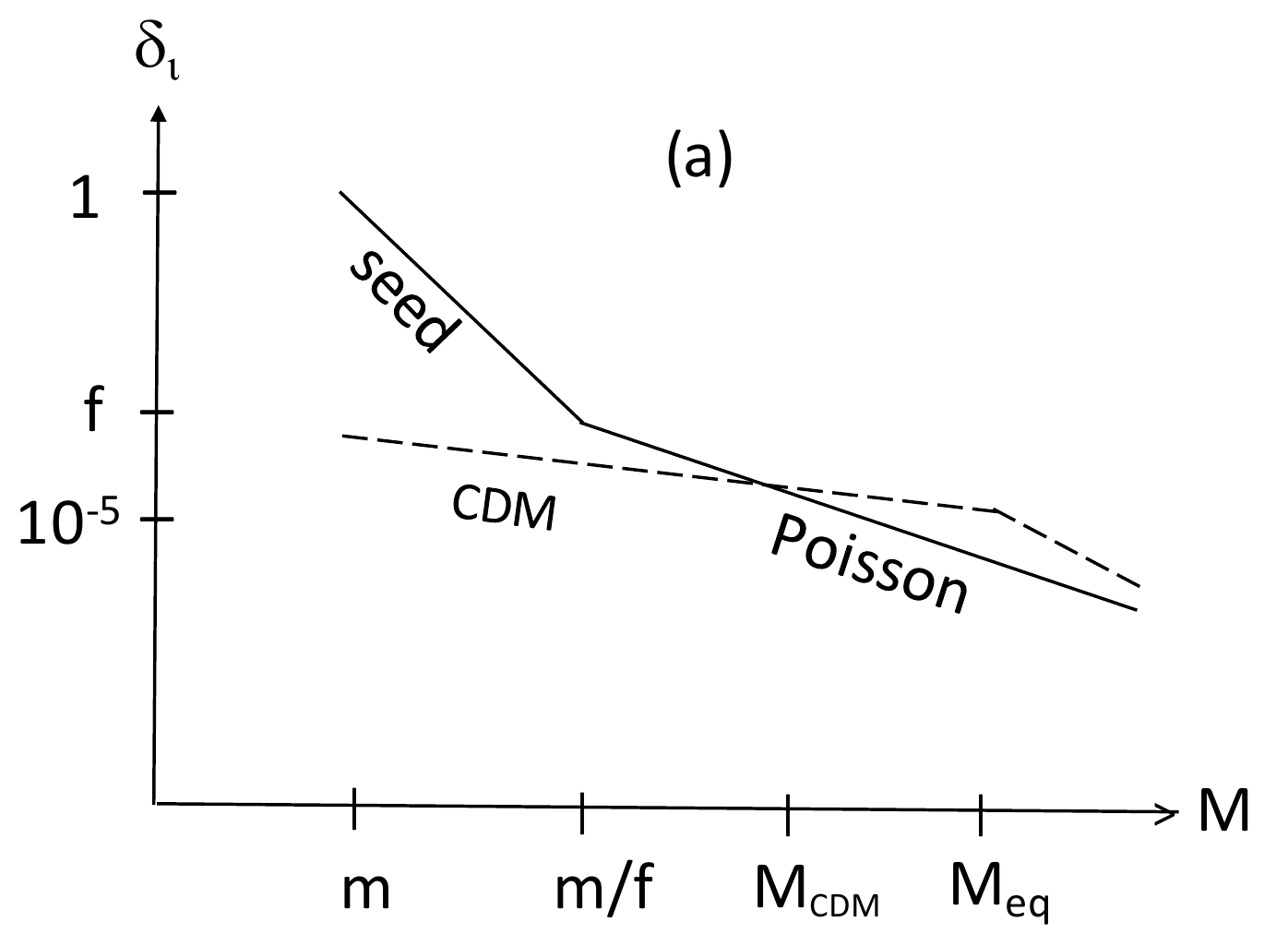,width=2.9in}
\psfig{file=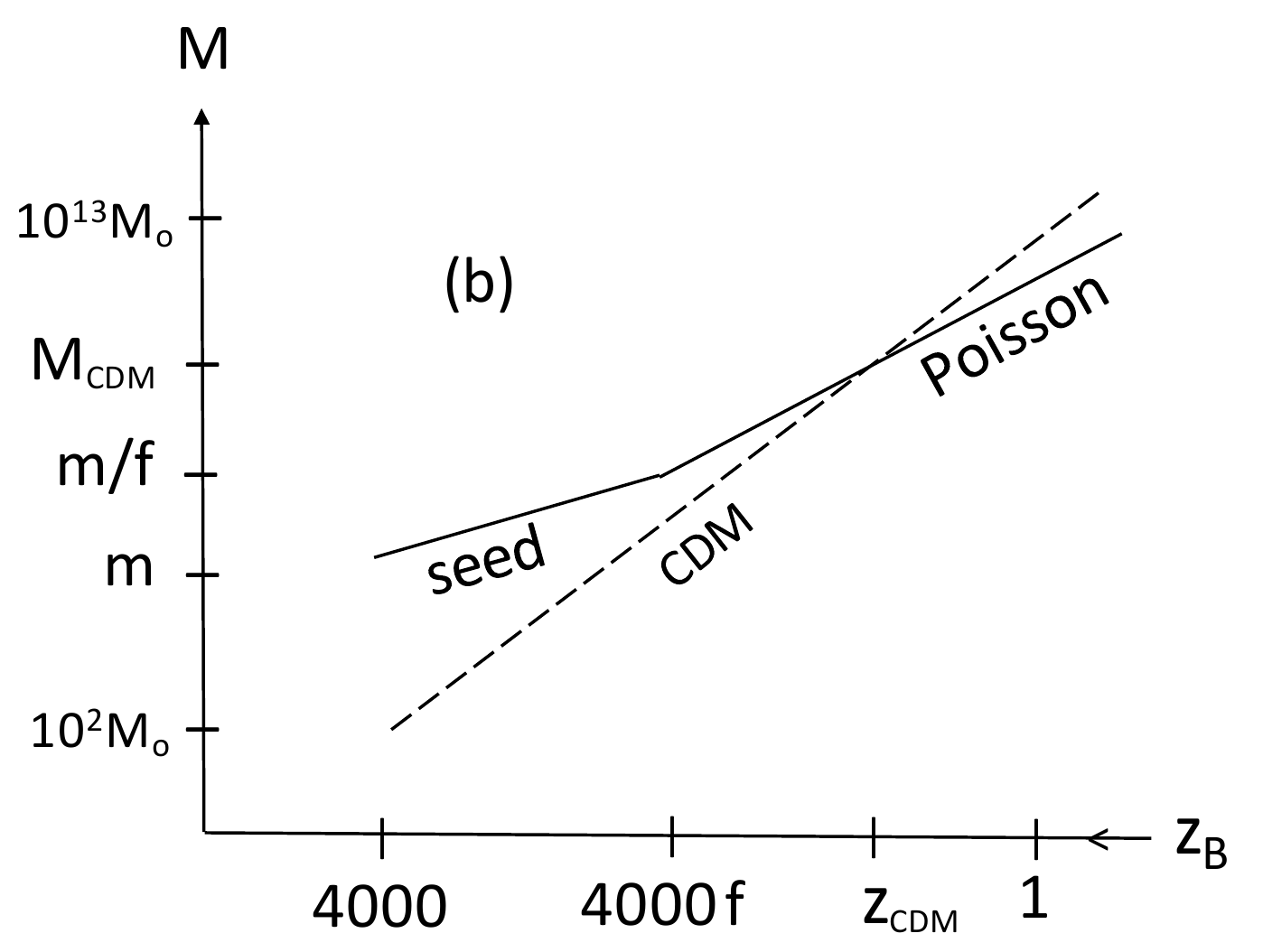,width=3.0in}
  \caption {(a) Form of initial fluctuation $\delta_i$ as a function of $M$ for the seed and Poisson effect with fixed $f$, the first dominating 
at small $M$ if $f $ is small but the second always dominating if $f \sim 1$. 
(b) Mass $M$ binding at redshift $z_B$ for fixed $f$, the Poisson effect dominating for low $z$ if $f$ is small  but at all $z$ if $f \sim 1$.
Also shown by dashed lines  are the forms for $\delta_i$ and $M(z_B)$ predicted by the CDM model, this indicating  the range $M > M_{CDM}$ and $z_B < z_{CDM}$ for which PBH fluctuations dominate.}
%(c) Maximum mass $M$ binding at redshift $z_B$ for values of $m$ allowed by wide-binary constaint on $f(m)$.}
 \label{fig2}
 \end{center}
 \end{figure}

In applying Eq.~(\ref{bind}),
we must first determine which effect dominates 
and this depends on the 
dark matter fraction. 
For a given value of $f$, Eq.~(\ref{bind}) and the condition 
$M<m/f$ imply that
the seed effect dominates for $z_B > z_{eq} f \sim 4000 f$. This condition is indicated in Fig.~\ref{fig2}(b)
and has a simple interpretation.  Since fluctuations grow as $(1+z)^{-1}$ after $z_{eq}$, the fraction of the Universe in gravitationally bound regions at redshift $z_B$ is $f z_{eq}/z_B$ and this exceeds $1$ for $f > z_B/z_{eq}$. In this case, competition between the seeds will  reduce the mass of each bound region to at most $M \sim m/f$, which is precisely the value of $M$ above which the Poisson effect dominates. So although the expression in Eq.~(\ref{bind}) for the mass bound by a seed  has no explicit dependence on $f$, there is an upper limit for its validity which does depend on $f$. On the other hand, the fraction of the Universe bound by the seeds is very small for $f \ll z_B/z_{eq}$, 
in which case the seed effect cannot berelevant for {\it most} cosmic structures. Indeed, it could play a dominant role in the formation of cosmic structures at the present epoch only for $f \sim z_{eq}^{-1}$, which would require fine-tuning. For larger $f$, the Poisson effect dominates; for smaller $f$, neither effect is  important.

If $f$ is is treated as a free parameter, unconstrained by observations, the dependence of $M$ on the redshift $z_B$ is as indicated in Fig.~\ref{fig2}(b). However, it is interesting to obtain the constraints on the function $M(z_B)$  implied by the limits  on $f(m)$ discussed in Sec.~II. 
If the PBHs provide the dark matter ($f\sim 1$), the Poisson effect always dominates and Eq.~(\ref{bind}) and the condition $m < 10^2M_{\odot}$ imply $M  < 10^{9} M_{\odot}$, which is much smaller than a galaxy. On the other hand, the mass bound by the seed effect can be as large as $M \sim z_{eq} m$ for $ f < z_{eq}^{-1}$, 
which can be in the galactic range for supermassive PBHs. 
However, unless one invokes  highly non-Gaussian fluctuations, 
the  $\mu$-distortion upper limit on $m$ of $10^5M_{\odot}$ 
still implies $M < 10^{9} M_{\odot}$. One can circumvent this limit if the PBHs grow appreciably  through accretion after $t_{eq}$ but the bound mass is then reduced because the Coulomb effect cannot operate for so long.

 One can constrain the mass bound by the seed and Poisson effects more precisely by considering specific limits shown in Fig.~\ref{limit}. For example, the wide-binary constraint (\ref{wb}), the condition $f <1$ and Eq.~(\ref{bind}) imply that the Poisson effect can bind a mass 
\begin{equation}
M < 
\begin{cases}
10^7 m z_B^{-2}
& ( m < 10^2\,M_\odot \quad {\rm or} \quad m > 10^7\,M_\odot )   \\
10^{9}z_B^{-2} M_{\odot} 
& (10^2\,M_\odot < m < 10^3\,M_\odot ) \\     
10^6 m z_B^{-2}
& (10^3\,M_\odot < m < 10^7\,M_\odot)\, , 
\end{cases}
\label{maxbind}
\end{equation}
with this dominating the seed effect for 
\begin{equation}
z_B < 
\begin{cases}
4000
&  ( m < 10^2\,M_\odot \quad {\rm or} \quad m > 10^3\,M_\odot ) \\
400 (m/10^3M_{\odot})^{-1} 
& (10^2\,M_\odot < m < 10^3\,M_\odot) \\
400
& (10^3\,M_\odot < m < 10^7\,M_\odot) \, .
\end{cases}
\end{equation}
%The combined constraints on $M(z_B)$ for different values $m$ are indicated in Fig.~\ref{fig2}(c).  
Of course, Fig.~\ref{limit} shows that there are numerous other constraints, so this one does not mean much on its own, but many of them have a similar form.

It is interesting to compare the seed and Poisson fluctuations with the primordial fluctuations implied by the CDM model. At the time of matter-radiation equality, $t_{eq} \sim 10^4$y, when the  PBH fluctuations start to grow, the CDM fluctuations have the form
\begin{equation}
\delta_{eq} \propto  
\begin{cases}
M^{-1/3}  
& (M < M_{eq}) \\
M^{-2/3}  
& (M > M_{eq}) \, ,
\end{cases}
\label{cdm}
\end{equation}
where $M_{eq} \sim 10^{16} M_{\odot}$ is the horizon mass at $t_{eq}$. These fluctuations and the effect on the binding mass are shown by the broken lines in Fig.~\ref{fig2}, the PBH effect being important whenever the solid line is above the broken line. This shows that there is generally  a mass $M_{CDM}$ below which the PBH fluctuation dominates, due either to the seed effect  for $f \ll 1$ or the Poisson effect for $f \sim 1$, so this produces  extra power on small scales. However, in the  mass range $M <M_{eq}$ relevant to the present considerations, the CDM fluctuations fall off slower than both the Poisson and seed fluctuations, so they generally dominate (i.e. the standard scenario is unchanged) for sufficiently large $M$. Whether both the seed and Poisson effect can be important at low $M$ depends on the precise normalisation of the CDM curve. In the mass range $M > M_{eq}$, 
the CDM fluctuations fall off slower than the seed fluctuation but faster than the Poisson fluctuation, so the latter could dominate again on very large scales, with CDM only dominating over some intermediate range of $M$.  However, this only applies on scales which are currently unbound. 
%As indicated in Fig.~\ref{fig2}, one therefore has two possibilities: either the seed and Poisson effects dominate on sufficiently small scales and the Poisson effect dominates again at very large (unbound) scales; or the seed effect dominates on small scales and the Poisson effect on (unbound) large scales but the CDM fluctuations dominate in between. 
There is also a redhsift $z_{CDM}$ below which the CDM fluctuations determine the binding mass.  

These effects have been invoked to produce three types of structure: the first bound baryonic objects (Kashlinksy 2016), the Lyman-$\alpha$ forest (Afshordi et al. 2003) and galaxies (Carr \& Silk 1983). If one has a monochromatic mass function, Fig.~\ref{fig2}(c) shows that one can only bind objects as large as galaxies if 
one invokes either highly non-Gaussian fluctuations or appreciable accretion after the $\mu$-distortion epoch. 
However, the PBH mass function is likely to be extended and one needs this anyway
 to produce an extended mass function for galaxies, so we now discuss this case.

\subsection{Extended PBH mass function}

If the PBHs have an extended mass function, both the seed and Poission effects could operate on different scales. Indeed, in principle, one could have two distinct PBH populations, both monochromatic but with a different mass. In this case, one population might provide the dark matter and generate a Poisson effect, while the other may provide a low density of SMBHs which generate a seed effect.  However, this seems rather contrived, so the following analysis assumes that the PBHs have a {\it continuous} mass spectrum. We first discuss the power-law case, since this is easiest to analyse and conveys the essential qualitative features. We will then consider the other possible mass functions described in Sec.~III.  

We first note that the competition between seeds can be neglected providing the fraction of the universe bound by them is small. In the power-law case, for seeds of mass $m$, this requires the mass of the bound regons to satisfy $M<  f(m)^{-1}m \propto m^{\alpha -1}$. Since this is an increasing function of $m$ for $\alpha >1$, we need to satisfy this requirement at the {\it lowest} value of $m$, leading to the condition $M < f_{dm}^{-1} m_{dm}$. In this case, each bound region  will contain a single seed and its mass will exceed that of the black hole by a factor  $z_{eq}/z_B \sim 10^3$, so the PBH mass spectrum will generate a galaxy mass spectrum of similar form. This is the simplest scenario 
but has the disadvantage that only some fraction of the gas goes into bound regions.

If the filling factor of the bound regions exceeds one, the situation is more complicated. In the monochromatic case, we saw that this corresponds to the Poisson effect becoming more important than the seed effect. In the extended case, it leads to a combination of the two effects. 
Since $f(m) \, m \propto m^{3 - \alpha}$, Eq.~(\ref{initial}) implies that the biggest $\sqrt{N}$ effect is associated with the largest holes providing $\alpha < 3$.
One expects this, for example, if the PBHs form from scale-invariant fluctuations 
when the equation of state is $p = \gamma \rho$ with $\gamma < 1 $. In this case,
 the dominant Poisson fluctuation on  scale $M$ is associated with the largest hole expected to be contained in such a region. Providing this is less than $m_{max}$, the effective value of $N$ is $1$, so the Poisson scenario reduces to the seed scenario with 
\begin{equation}
m_{seed}(M) =(f_{dm} M m_{dm}^{\alpha -2})^{1/(\alpha -1)} \quad (2 <\alpha <3) \, .
\label{seed}
\end{equation}
This necessarily increases with $M$ for $\alpha > 1$ and the mass binding at redshift $z_B$ is
\begin{equation}
M(z_B) \sim m_{dm} f_{dm}^{1/(\alpha -2)} (z_B/4000)^{(\alpha - 1)/(2 - \alpha)}  \,
\label{crit}
\end{equation}
from Eq.~(\ref{bind}), 
the exponent of $z_B$ exceeding the value $-1$ which applies in the monochromatic case. Thus, with an extended mass function, the seed mass is not fixed but depends on the mass of the region being bound. However, for $M$ sufficiently large that the mass given by Eq.~(\ref{seed}) exceeds $ m_{max}$, the dominant effect is the Poisson fluctuation associated with PBHs of mass $m_{max}$ and the number of them is
\begin{equation}
N_{max} \sim M m_{max}^{1-\alpha} m_{dm}^{\alpha -2}f_{dm}  \, .
\end{equation}
In this case, Eq.~(\ref{bind}) implies that the mass binding at redshift $z$ is
\begin{equation}
M(z_B) \sim m_{dm}^{\alpha -2}m_{max}^{3 - \alpha} f_{dm} (z_B/4000)^{-2} \, .
\label{crit2}
\end{equation}
From comparison with Eq.~(\ref{crit}), there is a change of slope at
\begin{equation}
z_*  \sim 4000  f_{dm} (m_{dm}/m_{max})^{\alpha -2} \quad (3 > \alpha > 2) \, ,
\end{equation}
this being the redshift below which the Poisson effect dominates, and the associated mass is 
\begin{equation} 
M_* \equiv M (z_*) \sim m_{dm}^{2 - \alpha} m_{max}^{\alpha -1} f_{dm}^{-1}  \, .
\label{newcrit}
\end{equation}
Note that 
$z_*$ decreases and $M_*$ increases as $m_{max}$ increases and $f_{dm}$ decreases. Indeed,
 the Poisson  effect dominates for all masses binding in the matter-dominated era ($z_B > 4000$) if
\begin{equation}
f_{dm} > (m_{dm}/m_{max})^{\alpha -2} \, .
\end{equation}
For a monochromatic mass function, $m_{dm} = m_{max} = m$ and $f_{dm} = f$, so Eq.~(\ref{newcrit}) just gives the mass $m/f$ which follows from Eq.~(\ref{initial}). 

If the PBHs form from scale-invariant fluctuations in the radiation era, $\alpha = 5/2$ and so Eq.~(\ref{crit}) implies that the mass binding from the seed effect at redshift $z_B$ is
\begin{equation}
 M(z_B) \sim 10^{12} z_B^{-3} f_{dm}^2 m_{dm} \, .
\end{equation}
For $f_{dm} \sim 1$ and $m_{dm} \sim 10 M_{\odot}$, this is of order a galactic mass for $z_B \sim 3$ and of order the mass of the first bound clouds for $z \sim 100$.
We discuss each of these cases in more detail later but numerical calculations would be needed to elucidate our treatment. 

This analysis can be extended to the other extended mass function scenarios. {In the {\it matter-domination} case, we can put $\alpha =2$ in the above analysis. However, many of the equations are invalid for $\alpha =2$ because the PBH density increases logarithmically with $m$, so most of the density is no longer in the smallest PBHs. One can consider either  a scenario with  $f_{dm} =1$ and the Poisson effect or a scenario with $f_{dm}  \ll 1$ and the seed effect.
In the {\it inflationary} case, the mass function is often given by Eq.~(\ref{mf}), so that it has a well defined peak but is broad and falls off relatively slowly either side of the peak. 
In the {\it critical collapse} case, the dominant contribution to both the dark matter density and the Poisson effect comes from the mass-scale $M_f$, so this is  like the monochromatic situation. 

 \subsection{Effect of compensating voids}

An important caveat is that one might expect each PBH seed to be initially surrounded by a compensating void, so there would be no excess mass to generate  a Coulomb effect. However, two processes would quickly remove this cancellation: (1) the black hole may escape the surrounding void due to the peculiar velocity generated by any asymmetry in its collapse (Fitchett 1983); (2) the void may escape the black hole by expanding until it is larger than the region being bound (Musco, private communication). This expansion is to be expected and if it occurs at the speed of light, it would have reached a radius of around $10$~kpc by the time of matter-radiation equality ($t_{eq} \sim 10^4$y), when the growth of the fluctuations is assumed to start. Coincidentally, this is comparable to the size of a galaxy.
Cai, Padilla \& Li (2015) have argued that the compensating void effect is always unimportant. 

\section{Constraints on $f(m)$ from formation of cosmic structures}

Even if PBHs do not play a role in generating cosmic sructures, one can still place interesting upper limits of the fraction of dark matter in them
by requiring that various types of structure do not form too early. In this section, we will consider the constraints associated with 
galaxies and clusters of galaxies.  We also consider the constraint associated with the first bound baryonic clouds, although this is not based on  direct observations but is merely the condition that the standard CDM prediction is not modified. We will then combine these constraints  into a single
constraint on the function $f(m)$. Throughout this section we assume that  the PBHs have a monochromatic mass function. How one can apply these limits for the case of an extended mass function is discussed, for example, by Carr, Kuhnel \& Sandstad (2016) and Carr et al. (2017a).

\subsection{Lyman-$\alpha$ forest}

Afshordi et al. (2003) used observations of the Lyman-$\alpha$ forest  
 to obtain an upper limit of about $10^4M_{\odot}$ on the mass of PBHs which provide the dark  matter. 
This conclusion was based on numerical simulations, in which the PBH Poisson effect  provides a flat contribution to the power spectrum $P(k)$ at sufficiently large wave-numbers.  
Carr et al. (2010) claimed to extend this result to the case in which the PBHs provide a fraction $f(m)$ of the dark matter.  However, there was some confusion in their argument, since they interpreted the Afshordi et al. limit as the requirement that Lyman-$\alpha$ clouds, assumed to have a mass  $M_{Ly \alpha} \sim 10^{10}M_{\odot}$, must not bind before $z_{Ly\alpha} \sim 4$ due to the Poisson effect. This led to a limit $f(m) < (m/10^4M_{\odot})^{-1}$.
However, the Afshordi at al. limit is independent of the nature of the  Lyman-$\alpha$ objects and just uses them as a tracer of the  power spectrum of the density fluctuations. Furthermore, most Lyman-$\alpha$ clouds are much smaller than $10^{10}M_{\odot}$. 
The proper extrapolation of their limit would still have roughly the form $f(m) < (m/10^4M_{\odot})^{-1}$ 
but only because it is indirectly related to the constraints  discussed below.
Indeed, the Carr et al.  requirement that $10^{10}M_{\odot}$ objects must not bind earlier than observed, while a misinterpration of the Afshordi et al. argument, is just a special case 
of the requirement that galaxies should not form too early.

\subsection{Galaxies and clusters}

In deriving the condition that galaxies do not form too early, it must be appreciated that galaxies span a wide range of masses and the upper limit on their formation redshift, $z_B(M)$, only refers to {\it typical} galaxies of mass $M$ (i.e. 
some galaxies of that mass  may form earlier than the average). Nevertheless, one can still obtain rough PBH constraints. For example, if we assume that Milky-Way-type galaxies have a mass of order $10^{12} M_\odot$ and must not bind before $z_B \sim 3$, we obtain 
\begin{equation}
f(m)
<
\begin{cases}
(m/10^6M_{\odot})^{-1}
& (10^6\,M_\odot < m \lesssim 10^9\,M_\odot ) \\
m/10^{12}M_{\odot}
& (10^9\,M_\odot \lesssim m < 10^{12}\,M_\odot)\ \, .
\end{cases}
\label{galaxy}
\end{equation}
This limit is shown in Fig.~\ref{fig3} and bottoms out at $m \sim 10^9M_{\odot}$ with a value $f \sim 0.001$. 
The first condition  in Eq.~(\ref{galaxy}) can  be obtained by putting $M \sim10^{12}M_{\odot}$ and $z_B \sim 3$ in Eq.~(\ref{bind}). The second condition corresponds to having just one PBH per galaxy
and is also the line above which  the seed effect dominates the Poisson effect ($f < m/M$). Indeed, since the initial seed fluctuation is $m/M$, the seed mass required for the galaxy to bind at $z \sim 3$ is immediately seen to be $10^9 M_{\odot}$.  
There is no constraint on PBHs below this line because the fraction of the Universe going into galaxies would be small, with most of the baryons presumably going into the intergalactic medium.
Therefore the seed effect does not modify the form of the limit shown in Fig.~\ref{fig3} but merely comes into play at the minimum. 

If we apply the same argument to dwarf galaxies, assuming these have $M \sim 10^{10} M_\odot$ and must  not bind before $z_B \sim 7$, we obtain 
\begin{equation}
f(m)
<
\begin{cases}
(m/5 \times 10^4M_{\odot})^{-1}
& (5 \times 10^4\,M_\odot < m \lesssim 2 \times10^{7}\,M_\odot )\ \\
m/10^{10 }M_{\odot}
& (2 \times 10^{7}\,M_\odot \lesssim m < 10^{10}\,M_\odot)\ \, ,
\end{cases}
\label{lyman2}
\end{equation}
this bottoming out at $m \sim 2 \times 10^{7}M_{\odot}$ with a value $f \sim 0.002$. 
On the other hand, if we apply the argument to clusters of galaxies, assuming these have a mass of $10^{14} M_\odot$ and must  not bind before $z_B \sim 1$, we obtain 
\begin{equation}
f(m)
<
\begin{cases}
(m/10^7M_{\odot})^{-1}
& (10^7\,M_\odot < m \lesssim 3 \times 10^{10}\,M_\odot )\ \\
m/10^{14 }M_{\odot}
& (3 \times 10^{10}\,M_\odot \lesssim m < 10^{14}\,M_\odot)\ \, ,
\end{cases}
\label{lyman2}
\end{equation}
this bottoming out at $m \sim 3 \times 10^{10}M_{\odot}$ with a value $f \sim 0.0003$. 

Although we are treating the various types of cosmic structures as distinct, it is clear that the above analysis can be applied to  bound structures of any mass. If  structures of mass $M$ are required to form after some redshift $z_B(M)$, the maximum value of $m$ for which $f \sim 1$ is allowed and the values of $m$ and $f$ where the constraint bottoms out are given by
\begin{equation}
\label{combined}
 m_{max} \sim 10^{-7} M z_B(M)^2, \quad m_{min} \sim 3 \times 10^{-4} M z_B(M),  \quad f_{min} \sim 3 \times 10^{-4} z_B(M) \, ,
\end{equation}
so we can merge the different limits  into a combined constraint, as indicated by the bold line in Fig.~\ref{fig3}. 
The limit scales as $m^{-1}$ at the low end,
as $m$ at the high end
and as some intermediate power of $m$ in between. In order to compare this constraint to the other limits on $f(m)$, it is also indicated by the bold broken line in Fig.~\ref{limit}. It is not as strong as the dynamical friction and accretion limits but these are more tentative. 

\subsection{First baryonic clouds}

The first baryonic clouds would be expected to have a mass of order $10^{6} M_\odot$ in the CDM picture. We cannot apply the above argument to  these directly because   
 there is no observational constraint on their formation redshift. However, we know that  the clouds would form at a redshift $z_B \sim 100$ in the CDM picture, so we can still derive a limit corresponding to the requirement that the standard picture is not perturbed.
This gives the condition:
\begin{equation}
f(m)
<
\begin{cases}
(m/10^3M_{\odot})^{-1}
& (10^3\,M_\odot < m < 3 \times 10^4\,M_\odot ) \\
m/10^{6}M_{\odot}
& (3 \times 10^4\,M_\odot < m < 10^{6}\,M_\odot)\ \, ,
\end{cases}
\label{lyman2}
\end{equation}
the limit bottoming out at $m \sim 3 \times 10^4M_{\odot}$ with a value $f \sim 0.03$. 
More generally, we must distinguish between direct observational constraints on the function $z_B(M)$, available for galaxies and clusters, and the form of the function {\it predicted} by some theory of structure formation}
The CDM scenario has nearly scale-invariant fluctuations at the horizon epoch and - as indicated by Eq.~(\ref{cdm}) - this implies that the density fluctuations at matter-radiation equality scale as $\delta_{eq} \propto M^{-1/3}$ 
in the mass range of interest. Then $ z_B \propto M^{-1/3}$, so Eq.~(\ref{combined}) implies
\begin{equation}
m_{min} \propto M z_B \propto z_B^{-2} \quad \Rightarrow\quad f_{min} \propto z_B \propto m_{min}^{-1/2} \, .
\end{equation}
The limit on $f(m)$ at low $m$, interpreted as the requirement that the CDM model remains unperturbed, is therefore as indicated by the broken line in Fig.~\ref{fig3}. That the theoretical  line on the left matches smoothly to the observational line on the right just reflects the fact that the CDM  model provides a good fit to the observations.
\begin{figure}
 \begin{center}
\psfig{file=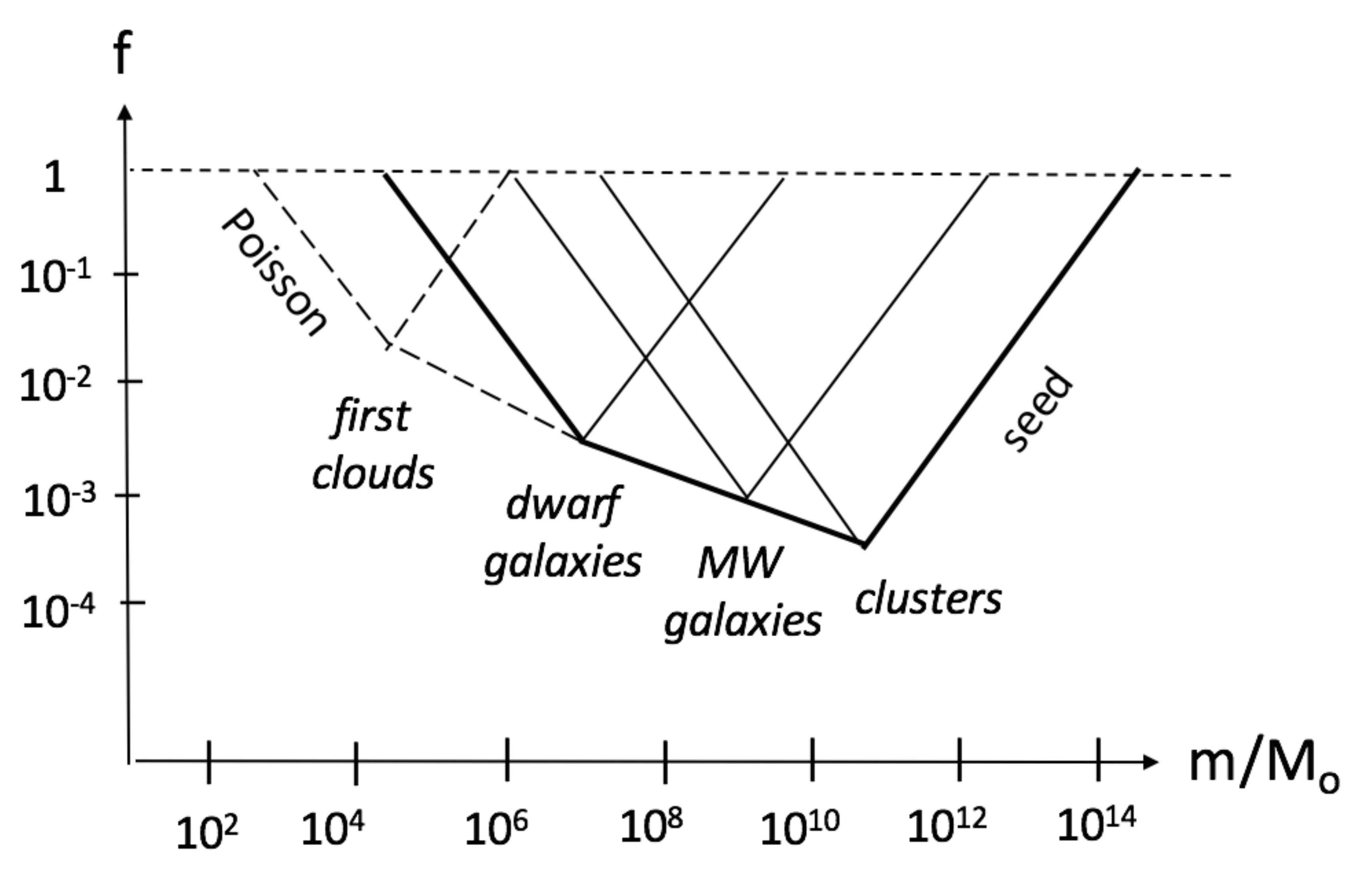,width=4.3in}
  \caption {Constraint on dark matter fraction $f$ in PBHs of mass $m$ from the requirement that the Poisson and seed fluctuations must not cause various types of cosmic structure to  form too early. The combination of the direct observational  constraints is indicated by the bold line; the broken lines on the left apply if one requires that the  standard CDM picture is not modified. }
 \label{fig3}
 \end{center}
 \end{figure}

\section{The role of PBHs in the formation of cosmic structure}
In the last section, we emphasised the constraints that can be placed on the number of large PBHs from the requirement that various types of objects do not form too early. In this section, we take a more positive approach, exploring the possibility that PBHs may have {\it helped} the formation of these objects, thereby complementing the standard CDM scenario of structure formation, a possibility also emphasized by Clesse \& Garcia-Bellido (2017a).
 
\subsection{Intermediate mass PBHs as seeds for SMBHs in galactic nuclei} 
There is clear evidence that SMBHs with mass $10^5 - 10^{10} M_{\odot}$ reside in the centres of most galaxies (Kormendy \& Richstone 1995; Magorrian et al. 1998; Richstone et al. 1998), with observations of  quasars  suggesting that these were already in place  at very early times ($z>6$). This includes the recent discovery of quasars powered by black holes of $1.2 \times 10^{10} M_{\odot}$ at $z=6.3$ (Wu et al. 2015) and  $8 \times 10^{8} M_{\odot}$ at $z=7.5$ (Banados et al. 2017). There is also the well-known correlation between by the mass of the SMBH and mass of the bulge, with the ratio being of order $10^4$ (Graham 2012; Sun et al. 2013; Reines \& Volonteri 2015), although  there may also be a correlation with the mass of the dark halo (Bogdan \& Goulding 2015).
The standard view is that these SMBHs formed through dynamical processes in galactic nuclei {\it after} galaxy formation (Rees 1984). There are then two possible pathways -- direct collapse to black holes (Habouzit et al. 2016) or super-Eddington growth (Pezzulli, Valiante \& Schneider 2016). Both of these have  been explored in a series of papers by Agarwal et al. (2012, 2013, 2014, 2016) but they are not without difficulties. In the former case,  
the  seeds are rare; in the latter case, they should be ultra-luminous and visible in deep X-ray surveys. 

There is also evidence that the linear relation between the SMBH and bulge mass  steepens at low mass (Graham \& Scott 2015), with several cases of central black holes in dwarf galaxies (Valluri et al. 2005). 
These are smaller than $10^6M_{\odot}$ and the 
$80$\% occupation number  found in nearby dwarfs suggests that 
they were of either Population III or primordial origin (Nguyen et al. 2017). There is also an ultra-compact dwarf galaxy of
$3 \times 10^8M_{\odot}$  containing a $2 \times 10^7M_{\odot}$ SMBH (Seth et al. 2014), possibly the stripped core of a previously massive galaxy (Ahn et al. 2015).  At high redshifts, SMBHs are predicted to be obese if early growth occurs (Agarwal et al. 2013) but current data are inconclusive (Shankar 2016).

It is therefore interesting to consider the possibility that quasars are seen at high redshift because they are powered by  SMBHs which formed {\it before} galaxies, In this case, they could be primordial (Bean \& Magueijo 2002; Duechting 2004; Khlopov et al. 2005; Clesse \& Garcia-Bellido 2015)
and this would lead to  three possible scenarios. \\

\noindent 
* The first possibility is that the PBHs were themselves  supermassive, so that they can be directly identified with the SMBHs. In this case, as discussed below, the black holes could also help to generate galaxies through either the seed or Poisson effect, the fluctuations growing by a factor of $4000$ between the time of matter-radiation equality and today.
This naturally expains the  proportionality between the black hole and galaxy  mass  and it could  provide an early mode of galaxy formation that might be important for the reionization of the universe (Chevallard et al. 2015). 
This would also have implications for 21cm observations of HI absorption in the dark ages, because of the longer path length of X-ray photons (Fialkov et al. 2017). \\

\noindent 
* The second possibility is that the PBHs had a more modest (intermediate) mass  and then grew through Eddington-limited accretion. This scenario was first suggested by Bean and Magueijo (2002), although they overestimated the amount of accretion in the early phase, and it has subsequently been advocated by other authors. 
%(Khlopov et al. 2005; Clesse \& Garcia-Bellido 2015). 
Bean and Magueijo argued that it needs a very narrow PBH mass function to reproduce the 
observed distribution of SMBHs, while Kawasaki, Kusenko \& Yanagida (2012)
suggested a specific inflationary scenario to account for this. 
However, most of the accretion still occurs after decoupling, so it may be difficult to distinguish this observationally from a scenario in which the black holes are non-primordial. In both cases,  one would expect a lot of radiation to have been generated and this may explain  part of the observed X-ray background (Soltan 1982). \\ 

\noindent 
* The third possibility is that the PBHs had a more modest mass and generated the SMBHs in galactic nuclei through the seed or Poisson  effect. 
For example, to produce a SMBH with $M \sim 10^{8}M_{\odot}$ by $z \sim 4$, the considerations of Sec.~\ref{sec4} show that one requires $m \sim 10^5M_{\odot}$ for the seed effect or $m \sim 10^2 M_{\odot}$ for the Poisson effect. However, the largest SMBHs have a mass $\sim 10^{10}M_{\odot}$ (Thomas et al. 2016), so in this case we would require $m \sim 10^7M_{\odot}$ for the seed effect or $m \sim 10^4 M_{\odot}$ for Poisson effect. Of course, one still has to explain how the bound region around an intermediate mass PBH  or a bound cluster of intermediate mass PBHs can evolve to a single SMBH.   
Accretion and merging could be important and only some fraction of the bound region may end up in the central black hole.\\
% in the Poisson case.

In the second and third
 scenarios, the SMBHs are not in place early enough for the galactic-scale fluctuations to experience the full growth factor of $10^4$, so galaxies would have to form from  primordial fluctuations in the usual way. Nevertheless, the presence of intermediate mass PBHs could still be advantageous 
in resolving various issues in dwarf galaxy formation (Silk 2017b),  
 especially since black hole recoils following binary mergers may suppress the presence of IMBHs in dwarfs that have undergone mergers as part of their formation history. Therefore one might not expect to end up with a single IMBH or SMBH at the centre  of the galaxy in the second and third scenarios. 
Recall also that one needs considerable accretion in order to avoid the $\mu$ constraint. Since the latter applies above $10^5M_{\odot}$, a PBH of final mass $m$ must increases its mass by at least a factor $m/10^5 M_{\odot}$.

\subsection{Supermassive PBHs as seeds for galaxies} 

We start with some historical remarks. Hoyle and Narlikar (1966) first suggested a version of the seed picture of galaxy formation in the context of the steady state theory. Their model starts with a fluctuation of the form $\delta = m /M$, like ours, but they are not constrained by the existence of a radiation-dominated era in selecting the time at which the fluctuation begins to grow. For reasons specific to the steady state model, they assumed that growth begins when the second-order term in $(GM/r)$ becomes comparable to the first-order term. 
Using a typical galactic mass  $M_{g} \sim 10^{11} M_{\odot}$, they required $m \sim 10^9 M_{\odot}$.
 Although the steady state theory is now superseded, Hoyle and Narlikar also discussed how deviations from spherical symmetry could give the range of shapes observed in elliptical galaxies, and how spirals could form from rotational effects. These features should apply in any seed theory.

Subsequently, Ryan (1972) also argued that SMBHs could seed galaxies.
Using a spherically symmetric Newtonian cosmology, he showed that the hydrodynamic equations permit a solution in which the density contrast has a particular form in the radiation-dominated and matter-dominated eras. This gave  expressions for the galactic mass $M_g \propto m^{2/5}$ and radius $R_g \propto m^{1/3}$. For our Galaxy, Ryan obtained $m \approx (1-9) \times 10^6 M_{\odot}$, which encompasses the now established mass of  $4 \times 10^6 M_{\odot}$ (Eckart \& Genzel 1996; Ghez et al. 1998). This analysis preceded the discovery of dark matter in galaxies and so is no longer applicable but it was still very prescient.

Gunn and Gott (1972) pointed out that one can make a very specific prediction about the structure of the galaxy resulting from the seed theory. If we assume that each shell of gas virializes after it has stopped expanding (i.e. settles down with a radius of about half its radius at maximum expansion), then one would expect the resultant galaxy to have a density profile $\rho (r) \propto r^{-9/4}$. This is because the shell with mass $M(r)$ binds at redshift
$z_B(r) \propto M(r)^{-1}$, with an associated radius $r \propto M(r)^{1/3}z_B(r)^{-1} \propto M(r)^{4/3}$ and density $\rho_B(r) \propto z_B(r)^3 \propto M(r)^{-3} \propto r^{-9/4}$.
This does not agree with the standard NFW profile (Navarro, Frenk \& White 1996), which goes from $r^{-1}$ at small radius to $r^{-2}$ at large radius, but one would not expect this to apply 
within the radius
of gravitational influence of the central black hole anyway (Gondolo \& Silk 1999). Indeed, 
hierarchical merging should rapidly erase memory of the profiles around the initial seeds, much as found in the highest resolution dark matter simulations (Angulo et al. 2016). 

We now turn to our own seed scenario for galaxy formation.
In this context,
we must first decide whether the PBHs have a monochromatic or extended mass function. In the monochromatic case, all the SMBHs in galactic nuclei would start off with the same mass and the galaxy mass just depends on the redshift at which it binds. Specifically, Eq.~(\ref{initial}) and the linear growth law $\delta \propto t^{2/3}$ for $t >t_{eq}$ imply 
that a mass $M$ binds at a time
\begin{equation}
t_B(M) \sim t_{eq} \left( \frac{M}{m} \right) ^{3/2} \sim 10^{10}   \left( \frac{M}{10^{12}M_{\odot}} \right) ^{3/2}   \left( \frac{m}{10^{8}M_{\odot}} \right) ^{-3/2} \mathrm{y} \, ,
\label{bind2}
\end{equation}
so  larger galaxies would form later, as in the standard CDM model. 
However, this scenario does not explain the observed correlation between the mass of the galaxy and the central black hole. This might still arise  if the black holes subsequently increase their mass through accretion to a value proportional to the galaxy mass. For example, the
effect of the galactic halo on the evolution of a central black hole has been discussed by Volonteri  
et al. (2011, 2016) and the observed  
dark-matter/SMBH ratio of $10^4$ might in principle be explained by Eddington-limited accretion (Silk \& Rees 1998). However, in this case Eq.~(\ref{bind2}) no longer applies since the  initial seed mass is much reduced. Therefore invoking a monochromatic PBH mass function does not seem very plausible.

For an extended PBH mass function, the PBH seeds will naturally produce a range of galactic masses at a given redshift. 
However, there are two distinct situations. In the first, the PBHs are sufficiently rare that there is only one per galaxy, which requires $f_{dm} < m_{dm}/M$, and the galaxy mass is then proportional to the seed mass.   
This naturally explains why the bulge mass is proportional to the  SMBH mass, with the ratio 
just  being the growth factor of fluctuations between the redshift of matter-radiation equality ($z_{eq} \sim 4000$) and the redshift when galaxies bind ($z_B \sim 3$).
One would also expect the galactic mass function to be the same as the PBH mass function. 
However, there should  be variances due to obese SMBHs in massive galaxies or anorexic IMBHs in dwarf galaxies.
While this scenario is the simplest, it only permits a small fraction of the Universe to go into galaxies. 

In the second situation,  the filling factor of the bound regions approaches $1$, so that the competition between seeds becomes important and one can no longer assume $M \propto m$. This situation is more complicated but there should still be a simple relation between the mass spectrum of the holes $dn/dm \propto m^{-\alpha}$ and that of the resulting galaxies. If $M_g \propto m^{\gamma}$, 
we expect the number of galaxies with mass in the range ($M, M + d M$) to be $dN_g(M) $ where
\begin{equation}
dN_g/dM \propto M^{(1 - \gamma -\alpha)/\gamma} \, .
\label{galspec}
\end{equation}
In this case, Eq.~(\ref{seed}) suggests $\gamma = \alpha -1$, so Eq.~(\ref{galspec}) gives 
\begin{equation}
dN_g/dM \propto M^{-2} \quad (M < M_{max}) \, .
\label{gal}
\end{equation}
This is independent of the value of $\alpha$ and converges to the usual equal mass per logarithmic mass interval limit.
The upper cut-off in $M$ in part reflects the value of $m_{max}$ but may also be determined by accretion effects, as discussed later. 
For comparison, the Schechter luminosity function is (Schechter 1976) 
\begin{equation}
\Phi (L) \propto L^{-1.07} \exp(-L/L_*) \, ,
\end{equation} 
while
he Press-Schechter mass function is (Press \& Schechter 1974)
\begin{equation}
dN_g/dM \propto M^{-2} \exp(-M/M_*) \, ,
\end{equation} 
with an exponential upper cut-off at  $M_* \sim 10^{12} M_{\odot}$. The integrated density $\rho_g(M) $ is then logarithmically divergent at the low mass end. 
It is striking that the observed mass function matches the 
prediction of Eq.~(\ref{gal}), providing one can explain 
 the exponential cut-off in some way.
The first scenario more naturally produces the proportionality between the black hole and galaxy mass but it only yields the Press-Schechter mass function if $\alpha \approx 2$.

We now consider the other
mass functions discussed in Sec.~III. Since the predicted galaxy mass function is independent of the value of $\alpha$, the above analysis should still apply in the matter-dominated scenario ($\alpha \approx 2$). 
The only difference is that the upper limit $M_{max}$ is now determined by the 
end of the matter-dominated epoch (Carr et al. 2017b).
One problem with a very extended PBH mass function is that the situation is dynamically complicated, with the larger PBHs  tending to sink to the centre of the galaxy through dynamical friction and then merging to form a single SMBH. Numerical simulations would be requred to ascertain even the qualitative features of this scenario.  We have seen that a lognormal mass-function is expected in some inflationary scenarios. In this case,  the spread of masses is much less, so the complications are reduced. For example, if the PBHs at the peak of the lognormal distribution provide the dark matter, then the ones on the high-mass tail would be sufficiently rare and massive to seed galaxies (Clesse \& Garcia-Bellido 2017c).
The PBHs would also  form clusters and this would have important consequences for their gravitational lensing effects (Clesse \& Garcia-Bellido 2017a). In the critical collapse case, the mass function is effectively monochromatic, since the PBH density falls off  very fast below the upper cut-off, so one cannot produce both the dark matter and seeds for galaxies.

\subsection{Upper limit on mass of galaxy seeded by PBH from accretion}

What determines the value of $M_{max}$ in Eq.~(\ref{gal})? Two factors impose an upper limit on the galactic mass which can be bound by a PBH seeed. The first is that if $M$ is larger than about $10^{12}M_{\odot}$, the bound region will not form a single galaxy but fragment into a  cluster of galaxies (Silk 1977). In this case, each galaxy would initially possess a central PBH seed but when the central galaxies merge to form a CD galaxy, the black holes might also merge to form a single SMBH. The proportionality between the galaxy and black hole mass might still pertain but clearly a more complicated dynamical analysis is required in this case.  
The second factor is accretion (Bower et al. 2016).
Accretion of the baryons cannot begin before decoupling because the Compton drag of the background photons prevents motion relative to the CMB. However, the baryons will quickly fall into the potential well created by the dark matter after decoupling and  accretion of dark matter will effectively begin from the time of matter-radiation equality ($t_{eq}$). 

During the radiation era, the sound-speed is $c_s \sim c$ and the accretion radius $R_a$ is just the Schwarschild radius, so the Bondi formula gives (Bondi 1952)
\begin{equation}
dm/dt \sim R_a^2 c_s \rho \sim (Gm^2)/(c^3 t^2) \, .
\end{equation}
Integrating this equation gives 
\begin{equation}
1/m - 1/m_i  \sim  (G/c^3) (1/t - 1/t_i)
\end{equation}
and hence 
\begin{equation}
	m \sim m_i/[1 - m_i/M_H(t_i)  + Gm_i/(c^3t)] \, .
\label{selfsimilar}
\end{equation}
Therefore there is very little accretion for $m_i \ll M_H(t_i)$ (i.e. for PBHs initially much smaller than the horizon). Although Eq.~(\ref{selfsimilar}) suggests $m \sim M_H(t)$ for $m_i \sim M_H(t_i)$, implying that a PBH with the horizon mass at formation should continue to grow like the horizon, this neglects the cosmic expansion. A more careful analysis shows that self-similar growth is impossible, so that accretion is negligible in the radiation era (Carr \& Hawking 1974).

During the matter-dominated era after $t_{eq}$, $R_a$ is increased (since $c_s $ falls below $c$) and so the accretion rate is also increased. Providing the matter temperature T follows the usual background evolution (i.e. neglecting  reheating), the Bondi formula  gives 
\begin{equation}
dm/dt \sim R_a^2 c_s \rho \sim (G^2m^2)/(G c_s^3 t^2) 
\sim Gm^2(kT_{eq}/m_p)^{-3/2}t_{eq}^{-2} \, .
\end{equation}
Integrating this gives 
\begin{equation}
1/m - 1/m_i  \sim  - \eta t \quad \mathrm{with} \quad  \eta \equiv G (m_p/kT_{eq}) ^{3/2} t_{eq}^{-2} \, .
\end{equation}
Hence 
\begin{equation}
m \sim m_i/(1-m_i \eta t) \, ,
\end{equation}
 which diverges at a time
\begin{equation}
\tau \sim 1/(\eta m_i) \sim (M_{eq}/m_i) (c_{eq}/c)^3 t_{eq} \, ,
\end{equation}
where 
$M_{eq} \sim c^3t_{eq}/G \sim 10^{16} \rm M_{\odot}$  is the horizon mass at $t_{eq} \sim 10^4$ y and $c_{eq} \sim c$. Thus the mass diverges at a time which precedes 
the present epoch ($t_o \sim 10^{10}$y) for
\begin{equation}
m_i > M_{eq} (t_{eq}/t_o) \sim 10^{10} M_{\odot} \, .
\end{equation}
This suggests that PBHs larger than $10^{10}  \rm M_{\odot}$ should not be found at the centres of galaxies because they would have swallowed the entire galaxy.
This argument complements recent ones of  Inayoshi \& Haiman (2016), who 
 find that  small-scale accretion physics and angular momentum transfer ultimately limits the SMBH mass to $1-6\times 10^{10}\rm M_\odot$, and Ichikawa \& Inayoshi (2017).  It is therefore interesting that observations at both low and high redshift  indicate a maximum SMBH mass of order  $10^{10} \rm M_{\odot}$.  Note that this argument assumes  accretion begins at $t_{eq}$ and therefore only applies if the black holes are primordial.

The above analysis assumes that the density and temperature at the accretion radius correspond to  the mean cosmological conditions. A more complicated analysis would be required if the growing bound cloud around the PBH ever became larger than the accretion radius.
Note also that the accretion rate reaches the Eddington limit when
\begin{equation}
dm/dt \sim \eta m^2 \sim m/t_{ED} \, ,
\end{equation}
where $t_{ED} \approx 4 \times 10^7$y is the Salpeter timescale (Salpeter 1964). However, we would only have super-Eddington accretion for
\begin{equation}
 m > (\eta t_{ED})^{-1} \sim M_{eq}(t_{eq}/t_{ED}) \sim 10^{12}M_{\odot} \, 
\end{equation}
and this never applies for the SMBHs of interest. 

\subsection{First clouds} 

Population III stars are made in the first clouds. They might also
be responsible for forming IMBHs in numbers that are marginally sufficient to seed the most massive SMBHs 
seen at $z \sim 6$, 
provided that super-Eddington accretion occurred (Pezzulli et al. 2017).
However, we have shown that one can also obtain 
a mass fraction of order $10^{-4}$ in intermediate mass PBHs (as required)
with conservative assumptions about the PBH mass function. In this case, one would not need 
the first clouds to collapse monolithically in order to form a population of IMBHs. At the very least, this seems to require rather special fine-tuning (Habouzit et al. 2016). 

Moreover, the mass of the first clouds is sufficiently small that the Poisson effect alone can bind them
if PBHs contribute sufficiently to the dark matter.
Let us consider the fiducial example of $100m_{100}\rm M_\odot$ PBHs contributing a fraction $f$ to the dark matter density. In the canonical LCDM scenario, Jeans mass fluctuations of mass  $10^6M_{J6} \rm M_\odot$ provide  the first DM-dominated dwarf galaxies at $z\sim 100$. These dwarfs, forming before reionization, are the building blocks of the next generation of dwarf galaxies, some of which may correspond to the extremely metal-poor ultra-faint dwarfs detected in recent deep surveys (Drlica-Wagner et al. 2015; Kim, Peter \& Hargis 2017; Newton et al. 2017).
The Poisson fluctuation imprint of PBHs  
on these scales  is 
$\sim 0.01 (fm_{100}/M_{J6})^{1/2}.$ 
This means that the first structures form at $z\sim  100 (f m_{100}/M_{J6})^{1/2}$, which may be 
earlier than in the usual scenario.  
For the fiducial parameters, one has to carefuly reexamine the limits from recombination due to Bondi accretion of gas onto the PBH, as discussed by Ricotti et al. (2008) and others. However, as discussed in Sec.~I,.
we can avoid this problem by considering a more conservative case in which  PBHs of mass $\sim 10^4\rm M_\odot$ are subdominant (eg. with $f\sim 0.01$). 

Let us now compare to the two scenarios. In the LCDM case, one can estimate the sizes, velocity dispersions and virial temperatures of the first systems as
 \begin{equation}
R \sim 40 M_{J6}^{1/3} \, \rm pc, \quad  
 \sigma\sim 10 M_{J6}^{1/3} \, \rm km/s, \quad
T\sim 10^4 M_{J6}^{2/3} \,\rm K \, .
\end{equation}
Residual ionization in these clouds leads to $H^{-}$ formation, eventually forming trace amounts of $H_2$ that allow cooling, fragmentation and formation of massive Population III stars. These short-lived stars generate  metallicity and pollute the IGM sufficiently to eventually lead to enhanced cooling and formation of dwarf galaxies. It is notoriously difficult to suppress fragmentation except in the vicinity of enhanced UV fields from neighbouring Population III star clusters
(Habouzit et al. 2016; Regan et al. 2017). There is inevitably competition between fragmentation and direct collapse, the trade-off 
involving the operation of disk instability (Inayoshi \& Haiman 2014). 
In the PBH scenario, the first cloud parameters are dramatically changed because of their boosted amplitude and earlier formation. They become 
\begin{equation}
R 
%\propto M^{1/3} z_B^{-1} 
\sim 600 \, f^{-1/2} M_{J6}^{5/6}m_{100}^{-1/2} \, \rm pc, \quad 
\sigma \sim  3 \, f^{1/4}M_{J6}^{1/12}m_{100}^{1/4} \,  \rm km/s, 
\quad T\sim 200 \,  f^{1/2}M_{J6}^{1/6}m_{100}^{1/2} \, \rm K \, .
\end{equation}
In this case, it seems likely that fragmentation is largely suppressed because of the lack of coolants and that runaway growth of the PBHs may ensue. Hence IMBH formation could precede the formation of the first dwarf galaxies. 
It is also possible that the first clouds are smaller than $10^6 M_{\odot}$ in the PBH scenario.
%  because the PBHs would increase the  gravitational binding. . 
In the usual LCDM scenario, halos smaller than this cannot  retain gas at the temperature $T\sim10^3$K expected due to $H_2$ cooling. However, the PBHs will enhance the trapping of gas in the gravitational potential of the first objects to form stars, hence lowering the critical mass.

Kashlinksy (2016) has also stressed that the Poisson fluctuations in PBH dark matter should lead to more abundant early collapsed halos than in the standard scenario.
He makes the interesting suggestion that the black
holes might generate the source-subtracted cosmic infrared background  fluctuations detected by the Spitzer and Akari satellites (Kashlinksy et al. 2005, 2007, 2012). These should correlate with the X-ray background fluctuations measured by Chandra and a recent paper suggests that this can be explained by accreting black holes of  possibly  primordial origin (Cappelluti et al. 2017).

\section{Gravity waves}

The proposal that the dark matter could comprise PBHs in the IMBH range has attracted much attention recently as a result of the LIGO detections of merging binary black holes with mass around $30\,M_{\odot}$  (Abbott et al. 2016a, 2016b, 2016c). Since the black holes are larger than initially expected, it has been suggested that they could represent a new  population.
One possibility is that they were of Population III origin (i.e. forming between decoupling and galaxies).  The suggestion that LIGO might detect gravitational waves from coalescing  intermediate mass  Population III black holes was first made by Bond \& Carr (1984)  more than 30 years ago and - rather remarkably - Kinugawa et al. (2014) predicted a Population III coalescence peak at $30 M_{\odot}$ shortly before the first LIGO detection.

Another possibility - more relevant to the considerations of the present paper and explored by many previous authors 
- is that the LIGO black holes were primordial.
 This does not necesarily require the PBHs to provide {\it all} the dark matter;  the predicted merger rate depends on too many uncertain astrophysical factors for the PBH number density to be specified precisely. However, several authors have made this connection (Clesse \& Garcia-Bellido 2015; Bird et al. 2016), with Clesse \& Garcia-Bellido (2017a) arguing that a lognormal distribution  centred at around $3 M_{\odot}$ naturally explains both the dark matter and the LIGO bursts without violating any of the current PBH constraints.  
On the other hand, others argue that the PBH density would need to be much less than the dark matter density to explain the LIGO results (Sasaki et al. 2016; Nakamura et al. 2016). Indeed, several groups have now used the LIGO results to constrain the PBH dark matter fraction (Raidal, Vaskonen \& Veermae 2017; Ali-Hamoud, Kovetz \& Kamionkowski 2017). Which alternative pertains depends on
whether the PBHs form binaries primordially or after they are clumped inside halos after galaxy formation, on whether they are uniformly distributed within halos or clustered (Clesse \& Garcia-Bellido 2017c) and on the PBH mass function. In the latter context, it should be stressed that the PBH density should peak at a lower mass than the coalescence signal for an extended PBH mass function, since the amplitude of the gravitational waves scales as the black hole mass. 

Although the origin of the 
the black holes associated with the LIGO events 
is still uncertain, 
future LIGO results and data from other gravitational-wave detectors -- such as eLISA (Seto 2016) and Pre-DECIGO (Kawamura et al. 2006)  -- might be able to distinguish between binary black holes of Population II, Population III or primordial origin. For example, Pre-DECIGO will be able to measure the mass spectrum and $z$-dependence of the merger rate. Another important clue may come from the spin distribution (Kocsis et al. 2017) and orbital eccentricities (Cholis et al. 2016) of the coalescing black holes.

As first stressed by Carr (1980), a population of massive PBHs would also be expected to generate a stochastic background of gravitational waves and this would be especially interesting if some of 
the PBHs were in  binaries  
coalescing due to gravitational radiation losses at the present epoch.  This was discussed by Bond \& Carr (1983) in the context of Population III black holes
and by Nakamura et al. (1997) and Ioka, Tanaka \& Nakamura (1998) in the context of PBHs.
Stochastic gravitational-wave backgrounds from black-hole binaries offer another way of distinguishing between the progenitors of the binaries.
Indeed, LIGO data had already placed weak constraints on the PBH scenarios a decade ago (Abbott et al. 2007) and an  updated analysis 
in the light of the recent merger events can be found in Abbott et al. (2016d), Dvorkin \& Barausse (2017) and Clesse \& Garcia-Bellido (2017c). 

If PBHs have an extended mass function, incorporating both dark matter at the low end and galactic seeds at the high end, this will have important implications for the predicted gravitational wave background. Theorists usually focus on  the gravitational waves  generated by either dark matter black holes (detectable by LIGO) or  
supermassive black holes in galactic nuclei (detectable LISA). However, with an extended PBH mass function, the gravitational wave background should encompass both these limits and also every intermediate frequency. This point has also been emphasized by Clesse \& Garcia-Bellido (2017b).  

\section{Discussion}

We have seen that PBHs in the intermediate to supermassive mass range could play several important cosmological roles. They could (1) explain the dark matter, (2) provide a source of LIGO coalescences, and (3) alleviate some of the problems associated with the CDM scenario
- including the formation of  SMBHs in galactic nuclei or even the first galaxies themselves.
Although the main focus of this paper has been (3), with particular emphasis on the seed or Poisson effect, 
 it is important to consider all three roles together. At one extreme, PBHs may play none of these roles, with our considerations merely  placing interesting constraints on the PBH scenario. At the other extreme, 
they may play all three. 

 Roles (1) and (2) are rather easily reconciled, since the mass scales involved are quite close and both in the IMBH range. Indeed, there is already a considerable literature on this topic. Reconciling (1) and (3) is more challenging, since the mass scales are very different, and this topic is relatively unexplored. Whether it is possible depends crucially on the PBH mass distribution. In principle, one could invoke two separate PBH populations and this might  conceivably arise  in some inflationary scenarios.
However, this seems less natural  than invoking a single PBH population,   in which case we need to distinguish between a monochromatic and extended mass function. 

For a monochromatic mass function,
if the PBHs provide {\it all} the dark matter ($f \sim 1$), then the Poisson effect dominates on all scales and various astrophysical constraints require $m < 10^2M_{\odot}$. This implies that PBHs can only bind subgalactic masses
but still  allows them to play a role in producing the first bound baryonic clouds
or the SMBHs which power quasars.
For $f \ll 1$, the seed effect dominates on small scales and can bind a region of up to $10^4$ times the PBH mass. 
However, if limits on
 the $\mu$-distortion in the CMB due to the dissipation of fluctuations before decoupling exclude PBHs larger than $10^5 M_{\odot}$, the seed effect may also only bind subgalactic scales.

If the PBHs have an extended mass function, they could both provide the dark matter and seed structure on the galactic scale. 
For a power-law mass function with $dn/dm \propto m^{-\alpha}$  up to some cut-off mass $m_{max}$, most of the mass is in the smallest PBHs 
for $\alpha >2$ (as expected) and the seed effect dominates below $M_* \sim m_{dm}^{2-\alpha}m_{max}^{\alpha -1}f_{dm}^{-1}$. However, 
this situation is dynamically complicated because of the large PBH mass range.
For a lognormal mass function, the PBH mass range is narrower, so the scenario is easier to understand and probably more plausible. For a critical mass function, most of the density is still concentrated at a single mass-scale, so the situation resembles the monochromatic one and PBHs cannot both provide the dark matter and galactic seeds. 

We stress that our proposal should be regarded as complementing rather than rivalling the CDM scenario, since there is no denying the success of the latter.
We also emphasize that our proposal has observational consequences that can be probed by future deep surveys in the optical, radio and X-ray frequency regimes.
In particular, 21cm dark-age experiments could play an important role in evaluating the contribution of PBHs  to early heating and/or ionization of the universe. Such experiments  could potentially discriminate between primordial and conventional sources of ionization  in the dark ages because of the differing redshift dependences and consequent implications for the hydrogen spin temperature evolution.

One aspect of our proposal which has been rather neglected in this paper is PBH accretion in the period after decoupling. It is clear that this is crucial in conventional scenarios for SMBH formation in galactic nuclei and it could be equally important in the primordial context. Indeed,  observations of accretion at $z > 100$ could be an interesting discriminant of the primordial  scenario, since this would not be expected in the Population III scenario. This also relates to the issue of whether the PBHs were initially in the intermediate or supermassive mass range. In order to generate cosmic structures without violating  the current $\mu$ constraints, we either need to invoke intermediate mass PBHs plus accretion or supermassive PBHs plus  high non-Gaussianity.

We should also comment on how our proposal helps to resolve some of the problems associated with the standard LCDM scenario. Many dwarf galaxy problems  can be solved by early feedback from a  central IMBH
when the dwarfs were gas-rich (Silk 2017b). This could  distinguish a PBH from an astrophysical IMBH formed by mergers. In the latter case, recoil is important, especially in shallower potential wells  (Choksi et al. 2017).
An especially important feature - that lacks alternative explanations - is the baryon deficit in massive galaxies. Supernova feedback fails to eject enough baryons  both for the present day Milky Way  and for massive dwarfs, the latter being a manifestation of the ``too big to fail'' problem (Garrison-Kimmel et al. 2013).
There are other solutions to this problem but these do not resolve the baryon deficit (Tomozeiu, Mayer \& Quinn 2016).
AGN feedback, most notably for the Milky Way, fails because the central SMBH is too small. 

The only solution may be to invoke IMBHs in dwarfs that are assembling to form galaxies like the Milky Way; in this case feedback may eject baryons well beyond the viral radius (Peirani et al. 2012).
The missing satellite problem is similarly resolved by IMBH feedback, although we note that the reality of this problem  has been questioned because of selection effects in counting the observed dwarfs (Kim, Peter \& Hargis 2017). Core  creation is facilitated by supernova feedback driving bulk gas motions that dynamically heat cusps (Pontzen \& Governato 2014) but AGN in dwarfs will produce similar effects. Indeed, this AGN-induced softening occurs  even in massive galaxies (Peirani et al. 2018). The formation of  a central SMBH from the mergers of IMBHs could result in the softening of the CDM cusp of the host galaxy
(Rashkov \& Madau 2014). Indeed, our own galaxy may have a kiloparsec-scale core rather than a cusp  (Portcail et al. 2017). 

Finally we comment on the implications of supermassive PBHs for primordial nucleosynthesis. It is sometimes claimed (Carr 2006) that the success of the BBNS scenario excludes PBHs forming after weak freeze-out, corresponding to initial PBH masses above $10^5\rm M_\odot$. However, this 
need not be true because at most $10^{-6}$ of the mass of the universe can be in PBHs at this time, even if they provide all the dark matter today. 
On the other hand, even a small fraction of  PBHs could have interesting consequences for primordial nucleosynthesis. For example, if we consider PBHs of $10^7 \rm M_\odot$ forming at $10^2$s, there should be an overproduction of  helium around each one because of the local density overenhancement, 
 The net cosmological effect will be at most $0.01\%$ because of the 
rarity of collapsed regions. Nevertheless,
if  mixing is inefficient, one might expect rare regions on dwarf-galaxy scales, optimistically 1 in $10^5,$ with  anomalous primordial nucleosynthesis abundances. 

\section*{Acknowledgments}
We thank A. Babul, S. Clesse, J. Garcia-Bellido, S. Khochfar, I. Musco, T. Nakama and M. Volonteri for useful discussions
We are also grateful to  Y. Sendouda for help in drawing one of the figures.
BC thanks Institut d'Astrophysique in Paris for hospitality received during this work. JS acknowledges the
support of the European Research Council via  grant  267117.


\begin{thebibliography}{99}
%\bibitem{abbott2007}
B. P. Abbott et al. (LIGO Collaboration), ApJ 659, 918 (2007).

\\%\bibitem{LIGO1}
B. P. Abbott et al. (Virgo, LIGO Scientific Collaboration), Phys. Rev. Lett. 116, 061102 (2016a).

\\%\bibitem{LIGO2}
B. P. Abbott et al. (Virgo, LIGO Scientific Collaboration), Phys. Rev. Lett. 116, 241103 (2016b).

\\%\bibitem{LIGO3}
B. P. Abbott et al. (Virgo, LIGO Scientific Collaboration), Phys. Rev. X. 6, 041015 (2016c).

\\%\bibitem{LIGO4}
B. P. Abbott et al. (Virgo, LIGO Scientific Collaboration), Phys. Rev. Lett. 116, 131102 (2016d).

\\%\bibitem{abit}
M.H.~Abitbol, J~Chluba, J.C.~Hill, B.R.~Johnson, MNRAS 471, 1126 (2017).
%arXiv:1705.01534 (2017).

\\%\bibitem{afshordi}
N.~Afshordi, P.~McDonald , D.N.~Spergel, ApJ Lett. 594, L71 (2003).

\\%\bibitem{agarwal1}
B.~Agarwal et al., MNRAS 425, 2854 (2012).

\\%\bibitem{agarwal2}
B.~Agarwal et al., MNRAS 432, 3438 (2013).

\\%\bibitem{agarwal3}
B.~Agarwal et al., MNRAS 443, 648 (2014).

\\%\bibitem{agarwal4}
B.~Agarwal et al., MNRAS 459, 4209 (2016).

\\%\bibitem{ahn}
C.P.~Ahn et al., ApJ 839, 72 (2017).

\\
C.~Alcock et al. (MACHO collaboration), ApJ Lett. 550, L169 (2001).
 
\\%\bibitem{ali}
Y.~Ali-Haimoud, M.~Kamionskoski, Phys. Rev. D 95, 043534 (2016).

\\%\bibitem{kovetz}
Y.~Ali-Hamoud, E.D.~Kovetz, M.~Kamionkowski, Phys. Rev. D 96, 123523 (2017). 

\\%\bibitem{angulo}
 R.E. Angulo, O. Hahn, A. Ludlow, S. Bonoli, MNRAS 471, 4687 (2016).
%arXiv:1604.03131 (2016). 

\\%\bibitem{banados}
E.~Banados et al., Nature 553, 473 (2018).
%arXiv:1712.01860 (2017).

\\%\bibitem{barrow}
J.D.~Barrow , P.~Coles, MNRAS  248, 52 (1991).

\\%\bibitem{bean}
R. Bean, J. Magueijo, Phys. Rev. D 66, 063505 (2002).

\\%\bibitem{bird}
S. Bird, I. Cholis, J.B. Muñoz, Y. Ali-Haïmoud, M. Kamionkowski, E. D. Kovetz, A. Raccanelli, A. G. Riess, Phys. Rev. Lett. 116, 201301 (2016).

\\%\bibitem{bogdan}
A.~Bogdan, A.D.~Goulding, ApJ 800, 124 (2015).

\\%\bibitem{bc}
J.R.~Bond, B.J.~Carr, MNRAS 207, 585 (1984).

\\%\bibitem{bondi}
H.~Bondi, MNRAS 112, 195 (1952).

\\%\bibitem{bower}
R.G.~Bower et al., MNRAS 465, 32 (2016).
%arXiv:1607.07445 (2016).

\\%\bibitem{brandt}
T. D.~Brandt, ApJ Lett. 824, L31 (2016).

\\%\bibitem{byrnes2012}
C.~Byrnes, E.~Copeland, A.~Green,  Phys. Rev. D. 86, 043512 (2012).

\\%\bibitem{cai}
Y.-C.~Cai, N.~Padilla, B.~Li, MNRAS 451, 1036 (2015).

\\
J.~Calcino, J.~Garcia-Bellido, T.M.~Davis, arxXiv:1803.09205 (2018).

\\%\bibitem{cappelluti}
N.~Cappelluti, R.~Arendt, A.~Kashlinsky, Y.~Li, G.~Hasinger, K.~Helgason, M.~Urry, P.~Natarajan, A.~Finoguenov, ApJ 847, L7 (2017).
%arXiv:1709.02824 (2017).

\\%\bibitem{carr1975}
B.J.~Carr, ApJ 201, 1 (1975).

\\%\bibitem{carr1977} 
B.J.~Carr, A\&A 56, 377 (1977).

\\%\bibitem{carr80}
B.J.~Carr,  A\&A 89, 6 (1980).

\\%\bibitem{carr81}
B.J.~Carr, MNRAS 194, 639 (1981).

\\%\bibitem{carr}
B.J. Carr, Inflating Horizons of Particle Physics and Cosmology, ed. H. Susuki et al, pp 129-149, Universal Academic Press (2006); arXiv: astro-ph/0511743. 

\\%\bibitem{ch1974}
B.J.~Carr, S.W.~Hawking, MNRAS 168, 399 (1974).

\\%\bibitem{carr-silk}
B. J.~Carr, J.~Silk, ApJ 268, 1 (1983).

\\%\bibitem{carr-lacey}
B. J.~Carr, C.G.~Lacey, ApJ 316, 23 (1987).

\\%\bibitem{carr-rees}
B.J.~Carr, M.J.~Rees. MNRAS 206, 801 (1984).

\\%\bibitem{cl}
B.J.~Carr, J.E.~Lidsey, Phys. Rev. D 48, 543 (1993).

\\%\bibitem{carsak}
B. J.~Carr, M.~Sakellariadou, ApJ 516, 195 (1999).

\\%\bibitem{cgl}
B.J.~Carr, J.H.~Gilbert, J.E.~Lidsey, Phys. Rev. D. 50, 4853 (1994).

\\%\bibitem{cksy}
B. J. Carr, K. Kohri, Y. Sendouda, J. Yokoyama, Phys. Rev. D 81, 104019 (2010).

\\%\bibitem{cks}
B.J.~Carr, F.~Kuhnel, M.~Sandstad,  Phys. Rev. D. 94, 083504 (2016).
%; arXiv:1607.06077.

\\%\bibitem{crtvv}
B.J.~Carr, M.~Raidal, T.~Tenkanen, V.~Vaskonen, H.~Veermae, Phys. Rev. D 96, 023514 (2017).

\\%\bibitem{ctv}
B.J.~Carr, T.~Tenkanen, V.~Vaskonen, Phys. Rev. D. 96, 063507 (2017).
%; arXiv:1706.03746.

\\%\bibitem{cksy2}
B. J. Carr, K. Kohri, Y. Sendouda, J. Yokoyama, preprint (2018a).

\\
B.J.~Carr, K.~Dimopoulos, C.~Owen, T.~Tenkanen, arXiv:1804.08639 (2018b).

\\%\bibitem{chapline}
G. F. Chapline, Nature (London) 253, 251 (1975). 

\\%\bibitem{chapfram}
G.F.~Chapline, P.~Frampton, JCAP 11, 042 (2016).

\\%\bibitem{chen}
L.~Chen, Q-G.~Huang, K.~Wang, JCAP 12, 044 (2016).
%arXiv:1608.02174 (2016).

\\%\bibitem{chevallard} 
J. Chevallard, J. Silk, T. Nishimichi, M. Habouzit, G.A, Mamon, S. Peirani, MNRAS 446, 3235 (2015). 

\\%\bibitem{chisholm}
J.R.~Chisholm, Phys. Rev. D. 73, 083504 (2006).

\\%\bibitem{chluba}
J.~Chluba, A.~Erickcek, I.~Ben-Dayan, ApJ 758, 76 (2012); arXiv:1203.2681.

\\
N.~Choksi et al., MNRAS 462, 1526 (2017). 
%arXiv:1707.06220 (2017).

\\%\bibitem{cholis}
I. Cholis, E.D. Kovetz, Y. Ali-Haïmoud, S. Bird, M. Kamionkowski, J.B. Muñoz, A. Raccanelli, Phys. Rev. D 94, 084013 (2016); arXiv:1606.07437.

\\M.W.~Choptuik, Phys. Rev. Lett. 70, 9 (1983).

\\%\bibitem{clesse}
S. Clesse, J. García-Bellido, Phys. Rev. D 92, 023524 (2015).
%; arXiv:1501.07565.

\\%\bibitem{cgb2017}
S. Clesse, J. Garcia-Bellido, arXiv:1711.10458 (2017a).

\\%\bibitem{clesse2}
S. Clesse, J. Garcia-Bellido, Phys. Dark Univ. 15, 142 (2017b).
%; arXiv:1603.05234.

\\%\bibitem{cgb2016}
S. Clesse, J. Garcia-Bellido, Phys. Dark Univ. 18, 105 (2017c).
%); arXiv:1610.08479.

\\%\bibitem{choptuik}
M.W.~Choptuik, Phys. Rev. Lett. 70, 9 (1993).

\\%\bibitem{vilenkin}
H.~Deng, A.~Vilenkin, arXiv:1710.02865 (2017).

\\%\bibitem{divalentino}
E. Di Valentino, E. Guisarma, M. Lattanzi, A. Melchiorri, O. Mena, Phys. Rev. D. 90, 043534 (2014).

\\%\bibitem{dolgov}
A. D. Dolgov, arXiv:1605.06749 (2016).

\\%\bibitem{ds}
A.~Dolgov, J.~Silk, Phys. Rev. D 47, 4244 (1993).

\\
A.~Drlica-Wagner et al., ApJ 813 109 (2015)
%.arXiv:1508.03622 (2015).

\\%\bibitem{dvorkin}
I.~Dvorkin, E.~Barausse, MNRAS 470, 454 (2017).

\\%\bibitem{duech}
N.~Duechting, Phys. Rev. D 70, 064015 (2004).

\\%\bibitem{eckart}
A.~Eckart, R.~Genzel, Nature 383, 45 (1996).

\\%\bibitem{EC}
C.R.\ Evans, J.S.\ Coleman, Phys.\ Rev.\ Lett.\ 72, 1782 (1994).

\\%\bibitem{fialkov} 
A. Fialkov, A. Cohen, R. Barkana, J. Silk, MNRAS 464, 3498 (2017). 

\\%\bibitem{fitch}
M.J.~Fitchett, MNRAS 203, 1049 (1983).

\\%\bibitem{frampton}
P. H. Frampton, Mod. Phys. Lett. A 31, 1650093 (2016).

\\%\bibitem{frametal}
P. H. Frampton, M. Kawasaki, F. Takahashi, T. T. Yanagida, JCAP,
%J. Cosmol. Astropart. Phys. 
04,  023 (2010).

\\%\bibitem{freese}
K.~Freese, R.~Price, D.N.~Schramm, ApJ 275, 405 (1983).

\\%\bibitem{gb2017}
J. Garcia-Bellido, J. Phys. Conf. Ser. 740, 012032 (2017).
% arXiv:1702.08275.

\\%\bibitem{clesse4}
J. Garcia-Bellido,, S. Clesse, arXiv:1710.04694 (2017).

\\%\bibitem{gb1996}
J.~Garcia-Bellido,  A.~Linde, D.~Wands, Phys. Rev. D. 54, 6040 (1996).

\\%\bibitem{morales1}
J.~Garcia-Bellido, E. Ruiz Morales, Phys. Dark  Univ. 18, 47 (2017).

\\
J.~Garcia-Bellido, M.~Peloso, C.~Unal, JCAP 9, 013 (2017).
%; arXiv:1707.02441.

\\%\bibitem{morales2}
J.~Garcia-Bellido, E. Ruiz Morales, Phys. Lett. B 776, 345 (2018).

\\
S.~Garrison-Kimmel, M.~Rocha, M.~Boylan-Kolchin, J.S.~Bullock, J.~Lally, MNRAS 433, 3539 (2013).

\\%\bibitem{ghez}
A.~Ghez et al., ApJ 509, 678 (1998). 

\\%\bibitem{gondolo}
P.~Gondolo, J.~Silk, Phys. Rev. Lett. 83, 1719 (1999).

\\%\bibitem{graham}
A.W.~Graham, ApJ 746, 113 (2012).

\\%\bibitem{grahamscott}
A.W.~Graham, N.~Scott, ApJ 798, 54 (2015).

\\%\bibitem{green}
A.M.~Green, Phys. Rev. D 94, 063530 (2016).

\\%\bibitem{gunn}
J.E.~Gunn, R.~Gott, ApJ 176, 1 (1972). 

\\%\bibitem{2016MNRAS.463..529H} 
M.~Habouzit, M~Volonteri, M.~Latif, Y.~Dubois, S.~Peirani, MNRAS 463, 529 (2016).

\\%\bibitem{habouzit} 
M. Habouzit, M.Volonteri, Y. Dubois, MNRAS 468, 3935 (2017). 

\\
C.~Hamadache et al., A\&A 454, 185 (2006).

\\%\bibitem{harada}
T. Harada,  C.M.~Yoo, K.~Kohri,  Phys. Rev. D 88, 08451 (2013).

\\%\bibitem{hyknj}
T. Harada, C.-M. Yoo, K. Kohri, K.-i. Nakao, S. Jhingan, ApJ 833, 61 (2016).
%; arXiv:1609.01588. 

\\%\bibitem{hz}
E.R.~Harrison, Phys. Rev. D. 1, 2726 (1971).

\\%\bibitem{horowitz}
B.~Horowitz, arXiv:1612:07264 (2016).

\\%\bibitem{hoyle}
F.~Hoyle, J.V.~Narlikar, Proc. R. Soc. London A 290, 177 (1966).

\\%\bibitem{zoltan}
K.~Ichikawa, K.~Inayoshi, arXiv:1704.00729 (2017).

\\%\bibitem{ih}
K.~Inayoshi, Z.~Haiman, MNRAS 445, 15491 (2014).

\\%\bibitem{ih}
K.~Inayoshi, Z.~Haiman, Ap J 828, 110 (2016).
%arXiv:1601.02511 (2016).

\\%\bibitem{inomata}
K.~Inomata, M.~Kawasaki, K.~Mukaida, Y.~Tada, T.-T.~Yanagida, Phys. Rev. D 95, 123510 (2017).

\\
Y.~Inoue, A.~Kusenko, JCAP 1710, 034 (2017).

\\%\bibitem{ioka}
K. Ioka, T. Tanaka, T. Nakamura, Phys. Rev. D 60, 083512 (1999).

\\%\bibitem{kmrv}
K.~Kannike, L.~Marzola, M.~Raidal , H.~Veermae, JCAP 09, 020 (2017).
%arXiv:1705.06225 (2017).

\\%\bibitem{kash}
A.~Kashlinsky, ApJ Lett.  823, L25 (2016). 

\\%\bibitem{kash2}
A.~Kashlinksy, R.G.~Arendt, J.C~Mather, S.H.~Moseley, Nature 438, 45 (2005).

\\%\bibitem{kash3}
A.~Kashlinksy, R.G.~Arendt, J.C~Mather , S.H.~Moseley, ApJ 654, L5 (2007).

\\%\bibitem{kash4}
A.~Kashlinksy, R.G.~Arendt, M.L.N.~Ashby, G.G.~Fazio, J.C~Mather, S.H.~Moseley, ApJ 753, 63 (2012).

\\%\bibitem{kawamura}
S. Kawamura et al., Class. Quant. Grav. 23, 2415 (2006).

\\%\bibitem{kky}
M.~Kawasaki, A.~Kusenko, T.~Yanagida, Phys. Lett. B. 711, 1 (2012).
%arXiv:1202.38482 (2012).

\\%\bibitem{kp}
M. Yu.~Khlopov, A.G.~Polnarev, Phys. Lett. B97, 383 (1980). 

\\%\bibitem{malomed}
M. Yu. Khlopov, B.A. Malomed, Ya.B. Zeldovich, MNRAS 215, 575 (1985).

\\%\bibitem{krs}
M. Yu.~Khlopov, S.G.~Rubin, A.S.~Sakharov, J. Astropart. Phys. 23, 265 (2005).

\\
S.Y.~Kim, A.H.G.~Peter, J.R.~Hargis,  et al. arXiv:1711.06267 (2017).

\\%\bibitem{kinugawa}
T.~Kinugawa, K.~Inayoshi, K.~Hotokezaka, D.~Nakauchi, T.~Nakamura, MNRAS 442, 2963 (2014).

\\%\bibitem{suyama}
B.~Kocsis, T.~Suyama, T. Tanaka, S.~Yokoyama, arXiv:1709.09007 (2017).

\\%\bibitem{kohri}
K.~Kohri, T.~Nakama, T.~Suyama, Phys. Rev. D. 90, 083514 (2014).
%; arXiv:1405.5999.

\\%\bibitem{Koike}
T.\ Koike, T.\ Hara, S.\ Adachi, Phys.\ Rev.\ Lett.\ 74, 5170 (1995).

\\%\bibitem{kormendy}
J. Kormendy, D. Richstone, ARA\&A 33, 581-624 (1995).

\\%\bibitem{kuhnel2016}
F.~Kuhnel, C.~Rampf, M.~Sandstad, Eur. Phys J. C76, 93 (2016).

\\%\bibitem{kous}
S.M.~Koushiappas, A.~Loeb (2017), Phys. Rev. Lett. 119, 041102 (2017).
%arXiv:1704.01668.

\\%\bibitem{ostriker}
C.G.~Lacey, J.P.~Ostriker, ApJ 299, 633 (1985).

\\%\bibitem{magorrian}
J.~Magorrian et al., AJ 115, 2285 (1998).

\\
E.~Mediavilla et al. Ap J 706, 1451 (2009).

\\%\bibitem{Meszaros2}
P.~Meszaros, A\&A 37, 225 (1974).

\\%\bibitem{Meszaros} 
P.~Meszaros, A\&A, 38, 5 (1975).

\\%\bibitem{metcalfe}
R.B.~Metcalf, J.~Silk, ApJ 464, 218 (1996).

\\%\bibitem{monroy}
M.A.~Monroy-Rodriguez, C.~Allen, ApJ 790, 159 (2014).
%, arXiv:1406.5169.

\\
B.~Moore, ApJ 413, L93 (1993).

\\%\bibitem{motohashi} 
H. Motohashi, W. Hu, Phys. Rev. D. 96, 063503 (2017).
%arXiv:1706.06784 (2017). 

\\%\bibitem{musco}
I. Musco,  J.C.~Miller, Class. Quant. Grav. 30, 145009 (2013). 

\\%\bibitem{nakama}
T.~Nakama, T.~Suyama, J.~Yokoyama, Phys. Rev. Lett. 113, 061302 (2014).

\\%\bibitem{nsy}
T.~Nakama, T.~Suyama, J.~Yokoyama, Phys. Rev. D 94, 103522 (2016).
%; arXiv:1609.02245.

%\\%\bibitem{nsk}
%T.~Nakama, J.~Silk, M. Kamionkowski, Phys. Rev. D 95, 043511 (2017); arXiv:1612.06264.

\\%\bibitem{ncs}
T.~Nakama, B.J.~Carr, J.~Silk, Phys. Rev. D 97, 043525 (2018).
%; arXiv:1710.06945.

\\%\bibitem{nakamura1}
T. Nakamura, M. Sasaki, T. Tanaka, K. S. Thorne, ApJ Lett. 487, L139 (1997).

\\%\bibitem{nakamura2}
T. Nakamura et al., arXiv:1607.00897 (2016).

\\%\bibitem{nfw}
J.F.~Navarro, C.S.~Frenk , S.D.M.~White, ApJ 462, 563 (1996).

\\
O.~Newton, M.~Cautun, A.~Jenkins, C.S.~Frenk, J.C.~Helly, arXiv:1708.04247 (2017).

\\%\bibitem{nguyen}
D.-D.~Nguyen, A.-C.~Seth, N.~Neumayer et al., arXiv:1711.04314 (2017).

\\%\bibitem{NJ}
J.C.\ Niemeyer, K.\ Jedamzik, Phys. Rev. Lett. 80, 5481 (1998).
%astro-ph/9709072.

\\
I.~Novikov, A.G.~Polnarev, A.A.~Starobinsky, Ya.B.~Zeldovich, A\&A 80,104 (1979).

\\
M.~Oguri, J.M.~Diego, N.~Kaiser, P.L.~Kelly, T.~Broadhurst, Phys. Rev. D. 92, 023518 (2018).
%arXiv:1710.001482 (2018).

\\
S.~Peirani, I.~Jung, J.~Silk, C.~Pichron, MNRAS 427, 2625 (2012).

\\
S.~Peirani et al., arXiv1801.09754 (2018).

 \\%\bibitem{pez} 
E. Pezzulli, R. Valiante, R. Schneider, MNRAS 458, 3047 (2016). 

\\%\bibitem{2017MNRAS.471..589P}
 E.~Pezzulli, M.~Volonteri, R.~Schneider, R.~Valiante, MNRAS 471, 589 (2017). 

\\%\bibitem{planck}
Planck collaboration. A\&A 594, A13 (2016).
%; arXiv:1502.01589.

\\%\bibitem{pk}
A.G.~Polnarev, M.Yu. Khlopov, Sov. Phys. Usp. 28, 213 (1985).

\\
A.~Pontzen, F.~Governato, Nature 506, 171 (2014).

\\
M.~Portail, O.~Gerhard, C.~Wegg, M.~Ness, MNRAS 465, 1621 (2017).

\\%\bibitem{poulin}
V.~Poulin, P.D.~Serpico, F.~Calore, S.~Clesse, K.~Kohri, Phys. Rev. D. 96, 083524 (2017).

\\%\bibitem{press-schechter}
W.H.~Press, P.~Schechter, ApJ 187, 425 (1974).

\\%\bibitem{quinn}
D. P. ~Quinn et al., MNRAS Lett. 396, L11 (2009).
%, 0903.1644.
 
\\%\bibitem{raidal}
M.~Raidal, V.~Vaskonen, H.~Veermae, JCAP 9, 037 (2017).

\\
V.~Rashkov, P.~Madau, ApJ 780, 187 (2014).

\\%\bibitem{rees}
M.J.~Rees, ARA\&A 22, 471 (1984).

\\
A.E.~Reines, M.~Volonteri, ApJ 813, 82 (2015).

\\
J.A.~Regan, E.~Vishal, J.H.~Wise, Z.~Haiman, P.J.~Johansson, G.L.~Bryan, Nature Astron. 1, 75 (2017).

\\%\bibitem{richstone}
D.~Richstone et al., Nature 383, A14 (1998).

\\%\bibitem{mor}
M.~Ricotti, J.P.~Ostriker, K.J.~Mack, ApJ 680, 829 (2008).

\\%\bibitem{ricotti}
M.~Ricotti, A.~Gould, ApJ 707, 979 (2009). 

\\%\bibitem{ryan}
M.P.~Ryan, ApJ Lett. 177, L79 (1972).

\\%\bibitem{salpeter}
E.~Salpeter, ApJ 140, 796 (1964).

\\%\bibitem{sasaki}
M. Sasaki, T. Suyama, T. Tanaka, S. Yokoyama, Phys. Rev. Lett. 117, 061101 (2016).

\\%\bibitem{schechter}
P.~Schechter, ApJ 203, 297  (1976).

\\%\bibitem{seth}
A.C.~Seth et al., Nature 513, 398 (2014).

\\%\bibitem{seto}
N. Seto, MNRAS 460, L1 (2016).

\\%\bibitem{shankar}
F.~Shankar, MNRAS 460, 3119 (2016).

\\%\bibitem{silk2}
J.~Silk, ApJ 211, 638 (1977)

\\%\bibitem{CDM}
J.~Silk, 14th International Symposium on Nuclei in the Cosmos (NIC 2016), 010101 (2017a).

\\%\bibitem{silk2017}
J.~Silk, ApJ Lett. 839, L13 (2017b).
 
\\%\bibitem{rees-silk}
J.~Silk, M.J.~Rees, A\&A 331, L1  (1998).

\\%\bibitem{soltan}
A.~Soltan, MNRAS  200, 115 (1982).

\\%\bibitem{sun}
A-L.~Sun et al. ApJ 778, 47 (2013).

\\%\bibitem{thomas}
J. Thomas, C-P. Ma, N.J.~McConnell, J.E. Greene, J.P. Blakeslee, R. Janish, Nature  532, 340 (2016)
 %arXiv:1604.01400.

\\
M.~Tomozeiu, l.~Mayer, T.~Quinn, ApJ 827L..15 (2016). 

\\%\bibitem{valluri}
M.~Valluri, L.~Ferrarese, D.~Merritt, C.L.~Joseph, ApJ  628, 137 (2005).

\\%\bibitem{vol1}
M.~Volonteri, P.~Natarajan, K.~Gultekin, ApJ 737, 50 (2011).

\\%\bibitem{vol2}
M.~Volonteri et al., The evolution of high-redshift massive black holes. IAU Symp. 319, 72 (2016); arXiv:1511.02588.
  %In {\it Galaxies at high redshift , their evolution over cosmic time} (2015).

\\
P.N.Wilkinson et al. , Phys Rev Lett 86, 584 (2001). 

\\%\bibitem{wu}
X-B.~Wu et al., Nature 518, 512 (2015).

\\%\bibitem{yoko}
J. Yokoyama, Phys. Rev. D 58, 083510 (1998).

\\%\bibitem{yoo}
J.~Yoo, J.~Chaname, A.~Gould, ApJ 601, 311 (2004).
%, astro-ph/0307437.

\\
M.~Zumalacarregui U.~Seljak, arXiv:1712.02240 (2017).


\end{thebibliography}
\end{document}